\documentclass[twocolumn,preprint]{aastex631}

\usepackage{xcolor}

\usepackage{natbib}
\usepackage{graphicx}
\usepackage{amsmath}
\usepackage{booktabs}
\usepackage{xspace}
\usepackage{hyperref}
\usepackage[T1]{fontenc}
\usepackage{lipsum}
\usepackage{comment}
\usepackage{multirow}
\usepackage{enumitem}

\shorttitle{Partial disruption of a planet around a white dwarf}
\shortauthors{Kurban et al.}

\begin{document}

\title{Partial disruption of a planet around a white dwarf: the effect of perturbation from the remnant planet on the accretion}

\correspondingauthor{Abdusattar Kurban}
\email{akurban@xao.ac.cn}
\correspondingauthor{Na Wang}
\email{na.wang@xao.ac.cn}
\correspondingauthor{Xia Zhou}
\email{zhouxia@xao.ac.cn}

\author[0000-0002-2162-0378]{Abdusattar Kurban}
\affil{Xinjiang Astronomical Observatory, Chinese Academy of Sciences, Urumqi 830011, Xinjiang, People's Republic of China}
\affil{Key Laboratory of Radio Astronomy, Chinese Academy of Sciences, Urumqi 830011, Xinjiang, People's Republic of China}
\affil{Xinjiang Key Laboratory of Radio Astrophysics, Urumqi 830011, Xinjiang, People's Republic of China}

\author[0000-0003-4686-5977]{Xia Zhou}
\affil{Xinjiang Astronomical Observatory, Chinese Academy of Sciences, Urumqi 830011, Xinjiang, People's Republic of China}
\affil{Key Laboratory of Radio Astronomy, Chinese Academy of Sciences, Urumqi 830011, Xinjiang, People's Republic of China}
\affil{Xinjiang Key Laboratory of Radio Astrophysics, Urumqi 830011, Xinjiang, People's Republic of China}

\author[0000-0002-9786-8548]{Na Wang}
\affil{Xinjiang Astronomical Observatory, Chinese Academy of Sciences, Urumqi 830011, Xinjiang, People's Republic of China}
\affil{Key Laboratory of Radio Astronomy, Chinese Academy of Sciences, Urumqi 830011, Xinjiang, People's Republic of China}
\affil{Xinjiang Key Laboratory of Radio Astrophysics, Urumqi 830011, Xinjiang, People's Republic of China}

\author[0000-0001-7199-2906]{Yong-Feng Huang}
\affil{School of Astronomy and Space Science, Nanjing University, Nanjing 210023, People's Republic of China}
\affiliation{Key Laboratory of Modern Astronomy and Astrophysics (Nanjing University), Ministry of Education, People's Republic of China}
\affiliation{Xinjiang Astronomical Observatory, Chinese Academy of Sciences, Urumqi 830011, Xinjiang, People's Republic of China}

\author[0000-0002-9061-6022]{Yu-Bin Wang}
\affil{School of Physics and Electronic Engineering, Sichuan University of Science \& Engineering, Zigong 643000, People's Republic of China}

\author[0000-0001-9227-3716]{Nurimangul Nurmamat}
\affil{School of Astronomy and Space Science, Nanjing University, Nanjing 210023, People's Republic of China}


\begin{abstract}
About 25\% -50\% of white dwarfs (WDs) are found to be polluted by heavy
elements. It has been argued that the pollution could be caused by
the tidal disruption of an approaching planet around the WD,
during which a large number of clumps would be produced and would
finally fall onto the WD. The reason that the planet
approaches the WD is usually believed to be due to
gravitational perturbations from another distant planet or stellar
companion. However, the dynamics of the perturbation and the detailed
partial disruption process are still poorly understood. In this study,
we present an in-depth investigation of these issues. A triple system
composed of a WD, an inner orbit planet, and an outer orbit
planet is considered. The inner plant would be partially disrupted
periodically in the long-term evolution. Fragments generated in the
process are affected by the gravitational perturbations from the
remnant planet, facilitating their falling toward the WD.
The mass loss rate of the inner planet depends on both its internal
structure and also on the orbital configuration of the planetary system.
\end{abstract}




\section{Introduction} \label{sec:intro}

White dwarfs (WDs) are the final product of the evolution of most stars (90\%) in
the Universe, providing important information for understanding the formation
and evolution of stars. During the evolution of WDs, heavy elements sink
into the interior under strong gravity, while light elements float up to the surface
so that there should be no heavy elements on the surface of the WDs \citep{Paquette1986ApJS}.
However, about 25\%--50\% of observed WDs have
strong emission or absorption lines of metals in their spectra, which points to the
presence of metal pollution in their atmospheres \citep{Zuckerman2003ApJ,Koester2014AA}.
The metal elements \citep{Veras2021orel.bookE,Klein2021ApJ,Budaj2022AA} that pollute
the atmosphere of WDs may originate from disintegrated asteroids or
planets \citep{Debes2002ApJ,Jura2003ApJ,Antoniadou2016MNRAS,Veras2019MNRAS.488..153V,
    Duvvuri2020ApJ...893..166D,Veras2020MNRAS.493.4692V}.
In a WD-asteroid/planet system, if the asteroid or planet enters the tidal
disruption radius of the central object, it will be disintegrated under the action
of tidal forces, forming a ring or disc \citep{Veras2014MNRAS,Malamud2020a}. The
generated fragments could be accreted onto the surface of the WD under
various interactions.

According to the theory of stellar evolution, the progenitor of a WD would
significantly expand during the post-main sequence evolution stages and engulf any
planets that previously existed within a few au ($1.5\times10^{13}$ cm) around
the host star \citep{Sato2008PASJ,Villaver2009ApJ,Kunitomo2011ApJ}. However,
increasing evidence suggests that WDs could have close-in celestial companions,
which include multiple fragments or
asteroids \citep{Granvik2016Natur,Xu2016ApJl,Vanderbosch2021ApJ,Farihi2022MNRAS},
debris discs/rings \citep[e.g.,][]{Koester2014AA,Farihi2016NewAR}, minor
planets \citep{Vanderburg2015Natur,Manser2019Sci,Blackman2021Natur,Vanderbosch2020ApJ,Guidry2021ApJ},
and even major planets \citep{Thorsett1993ApJ,Luhman2011ApJ,Gansicke2019Natur,Vanderburg2020Natur}.

The origin of these close-in asteroids and planets is still highly uncertain,
but intuitive speculation is that they might be the outcome of various dynamic processes.
Firstly, scattering in a multi-planet system can form a close-in planet around the WD.
Strong short-term gravitational interactions in the multi-planet system cause orbital
instabilities, i.e., one planet may be perturbed by another planet with approximately
equal or greater mass and would be eventually scattered toward the WD
\citep{Debes2002ApJ,Bonsor2011MNRAS,Debes2012ApJ,Veras2013MNRAS.431.1686V,
    Frewen2014MNRAS,Bonsor2015MNRAS,Veras2016MNRAS.458.3942V,Veras2017MNRAS.465.2053V,
    Veras2018MNRAS.481.2180V,Mustill2018MNRAS}.
The gravitational instability increases for a smaller planet mass, a larger orbital
eccentricity, and the effect is enhanced when the number of planets increases in
the system \citep{Maldonado2020aMNRAS,Maldonado2020bMNRAS,Maldonado2021MNRAS,Veras2021orel.bookE}.
Such gravitational instabilities in multi-planet systems increase metal pollution
of WDs and can explain the observed fraction of polluted WDs
\citep{Li2021MNRAS,Maldonado2022MNRAS,Stock2022MNRAS,Connor2022MNRAS}.
Secondly, it has been argued that secular effects such as the Kozai-Lidov
mechanism \citep{Lidov1962,Kozai1962,Naoz2016ARAA}
might be a feasible approach to form close-in planets around WDs. The perturbation from a distant planet or
stellar companion can push a planet to the tidal disruption region around the WD
\citep{Bonsor2015MNRAS,Hamers2016MNRAS,Petrovich2017ApJ,Diego2020ApJ,Stephan2017ApJ,Stephan2021ApJ}.
Thirdly, the capture of a free-floating planet by a WD \citep[e.g.,][]{Kremer2019ApJ} or
WD planetary system \citep[e.g.,][]{Goulinski2018MNRAS}, or the exchange of a planet between
the binary stars \citep[e.g.,][]{Kratter2012ApJ...753...91K} can also form a close-in planetary system.
The close-in planet WD J0914+1914 b is most likely the outcome of scattering
\citep{Gansicke2019Natur, Veras2020MNRAS.492.6059V} or the Kozai-Lidov migration
\citep{Stephan2021ApJ}, while the formation of WD 1856+534 b might be ascribed to the
Kozai-Lidov effect \citep{Diego2020ApJ, Connor2021MNRAS}.

To understand the metal pollution in the atmosphere of WDs, it is
necessary to study the accretion process of disintegrated
fragments. Previously, the process has been investigated
in context of various mechanisms such as the
Poynting-Robertson drag effect
\citep{Rafikov2011ApJ,Veras2014MNRAS,Veras2015MNRAS.451.3453V,Veras2020MNRAS.493.4692V},
the collisional grind down
\citep{Jura2003ApJ,Wyatt2007ApJ,Li2021MNRAS,Brouwers2022MNRAS},
the Yarkovsky effect
\citep{Rafikov2011ApJ,Veras2014MNRAS,Veras2015MNRAS.451.3453V,Veras2020MNRAS.493.4692V},
the gravitational perturbation of a distant planet on fragments
\citep{Li2021MNRAS}, the action of previously existing matter
\citep{Malamud2021MNRAS}, the Alfv\'{e}n wave drag caused by
magnetic field \citep{Hogg2021MNRAS,Zhang2021ApJ}, and sublimation
\citep{Veras2015MNRAS.452.1945V}. Some authors \citep{Li2021MNRAS,
Brouwers2022MNRAS, Veras2022MNRAS} have even considered the joint
effect of multiple mechanisms mentioned above to analyze the
dynamic process of debris accretion. In these works, the
accretion timescale is usually longer than $ 10^{4}$ yr and the
accretion rate in the asteroid disruption scenario is typically
smaller than $10^{13}$ g s$^{-1}$.

As previously mentioned, the processes such as scattering, capturing,
and the Kozai-Lidov mechanism can lead to highly eccentric, close-in
planetary systems. In these scenarios, the formation of a close-in orbit
may occur gradually, depending on the configuration of the planetary system:
    \begin{itemize}
    \item The gradual approach occurs under the action of the Kozai-Lidov mechanism
          alone. For a planet in a stable three-body system where dynamical
          instability never occurs, the long-term Kozai-Lidov effect makes it
          possible that the planet's periastron distance gradually decreases.
    \item The gradual approach could occur under the combined effect of the
          short-term dynamical processes (scattering or capturing) and the
          long-term Kozai-Lidov effect. For the combination of scattering
          and the Kozai-Lidov effect, stable eccentric orbit configuration
          with periastron distance lying outside the partial disruption radius
          forms first due to the planet-planet scattering, then the planet
          in the inner orbit further evolves under the Kozai-Lidov perturbations
          from the outer giant planets \citep{Nagasawa2011ApJ} or binary
          companion \citep{Mustill2022MNRAS}, causing a gradual decrease
          of the periastron distance. For the combination of capturing and
          Kozai-Lidov effect, stable eccentric orbit configuration forms via
          the capture of a planet by a WD-planet
          system \citep[e.g.,][]{Goulinski2018MNRAS, Kremer2019ApJ} first,
          then the Kozai-Lidov effect due to the outer planet further excites
          the eccentricity of the inner orbit planet in subsequent evolution
          process, which results in a gradual decrease in the periastron distance.
    \end{itemize}
The planet may experience many partial disruptions during the long-term
approaching process and the main portion of the planet could survive after
each partial disruption. In this study, we investigate the accretion of tidal
debris by the WD. The gravitational perturbation from the remnant
planet on the accretion will be considered.

The structure of our paper is organized as follows.
In Section \ref{sec:framework}, we describe the model used to analyze the orbit evolution of a triple system.
In Section \ref{sec:mass-evolution}, the mass loss during the planet's long-term orbit evolution will be calculated.
The effects of gravitational perturbations from the remnant planet on the accretion of tidal fragments are studied in Section \ref{sec:accretion}.
The fate of the remnant planet and the actions of other forces that affect the tidal debris are discussed in Section \ref{sec:discussion}.
Finally, Section \ref{sec:conclusion} presents our conclusions.


\section{Model } \label{sec:framework}

\begin{figure*}[t!]
    \plotone{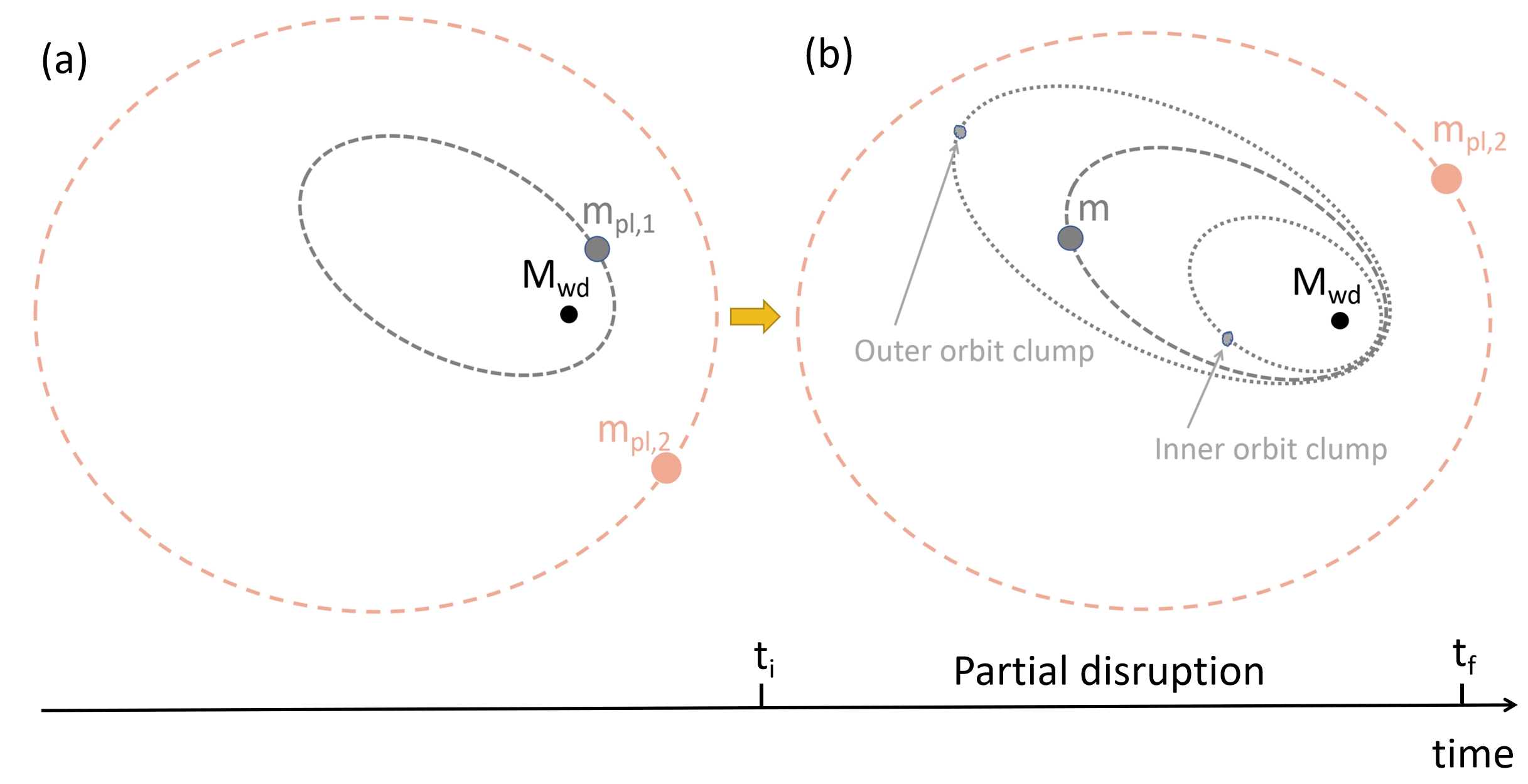}
    \caption{Schematic picture (not to scale) for the evolution of a triple system composed of a WD $M_{\rm WD}$, a planet $m_{\rm pl,1}$ in the inner orbit, and a planet $m_{\rm pl,2}$ in the outer orbit. (a) Orbital evolution prior to partial tidal disruption, where $m_{\rm pl,1}$ remains safe until its orbit evolves to the time point $t_{\rm i}$ at which the partial disruption starts. (b) Illustration of the partial disruption of the planet $m_{\rm pl,1}$, where $m$ represents the remnant of $m_{\rm pl,1}$ in each partial disruption that occurred during the time interval $\Delta t = t_{\rm f} - t_{\rm i}$, where $t_{\rm f}$ is the stopping time of partial disruption. The dotted ellipses represent the orbits of clumps in the inner and outer streams with respect to the orbit of $m$.}
    \label{fig:fig1}
\end{figure*}

The disruption of a planet in a multi-planet system can account for the WD pollution.
In this case, the eccentricity of the inner orbit planet must be excited to an extreme
value by the gravitational perturbations from the outer orbit objects. For simplicity,
let us consider a planetary system composed of a central WD and two planets
as illustrated in Figure \ref{fig:fig1}.
According to Kepler's third law, the orbital period ($ P_{\rm orb} $) and semi-major
axis ($ a $) of a planet are related to each other
as $P_{\rm orb}^2 = 4 \pi^2 a^3/\left[G(M_{\rm WD} + m_{\rm pl})\right]$,
where $m_{\rm pl}$ is the mass of the planet. The distance between the WD
and the planet at phase $ \theta $ is $r = a(1-e^2)/(1 + e \,\cos\theta)$ for an
eccentric orbit, where $ e $ is the orbital eccentricity. The argument and ascending
node of the orbit are denoted as $\omega$ and $\Omega$, respectively. We will use
subscripts 1 and 2 to describe the parameters of the inner and outer orbit planets,
respectively. The relative inclination angle between the two orbits is denoted as $i$.

\subsection{Structural parameters of planets}

Currently, the observed mass range of planets is very wide. It was speculated
that the low-mass (super-Earth or Neptune-like) planets may be equally common in
a wide range of orbits around the progenitors of WDs and contribute to some fraction of WD
pollution \citep{Veras2016MNRAS.458.3942V, Mustill2018MNRAS}. Recently, rocky planets of
masses $\sim 40 M_{\oplus}$  \citep[TOI-849 b,][]{Armstrong2020Natur}
and $\sim 73 M_{\oplus}$ \citep[TOI-1853 b,][]{Naponiello2023Natur} have been
observed, thought to be composed of a heavy element core and a thin
hydrogen-helium atmosphere. The dynamical evolutionary properties of such
low-mass planetary systems have been extensively investigated, paying special
attention to the mass distribution function \citep{Maldonado2020aMNRAS, Maldonado2020bMNRAS}. In this study,
we will investigate similar planetary systems composed of such planets.

\begin{figure}
    \plotone{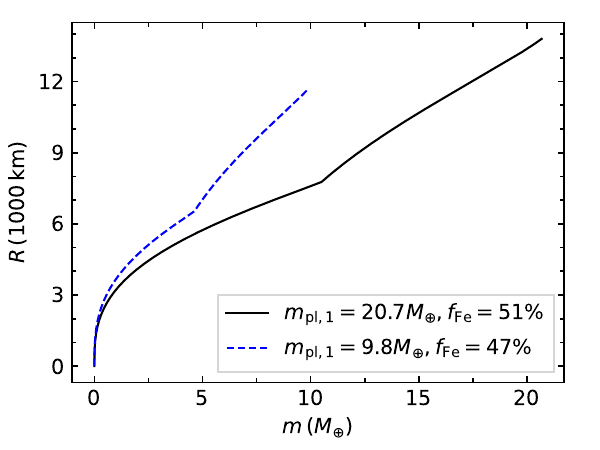}
    \caption{Internal mass and radius profile for a planet
        with $m_{\rm pl,1} = 9.8 M_{\oplus}$ (dashed line)  or
        $m_{\rm pl,1} = 20.7 M_{\oplus}$ (solid line).
    }
    \label{fig:fig2}
\end{figure}

For simplicity, a two-layer (iron core and rocky mantle) rocky
planet is assumed to be in the inner orbit. The main parameters
(mass and radius) of the planet are calculated by using the
equation of states (EOSs) widely used in exoplanet modeling,
including iron (Fe) material considered in \cite{Smith2018NatAs}
and rocky material (MgSO$_3$) considered in \cite{Seager2007ApJ}.
Therefore, we take the parameter sets as $m_{\rm pl,1} = 9.8
M_{\oplus}, 20.7 M_{\oplus}$, $m_{\rm pl,2} = 9.55\times10^{-4}
M_{\sun} (1 M_{\rm Jup})$, $M_{\rm WD} = 0.6 M_{\sun}$. Figure
\ref{fig:fig2} shows the internal mass-radius profile for the two
exemplar planets. Some key parameters (total mass $m_{\rm
pl,1}$ and radius $R_{\rm pl,1}$, iron core mass $m_{\rm
pl,1}^{\rm Fe}$ and radius $R_{\rm pl,1}^{\rm Fe}$, and iron mass
fraction $f_{\rm Fe}=m_{\rm pl,1}^{\rm Fe}/m_{\rm pl,1}$) are
listed in Table \ref{tab:table1}.

\begin{table}[h!]
    \centering
    \caption{Key parameters of two exemplar planets considered in this study.\label{tab:table1}}
    \begin{tabular}{lcccc}
        \hline\hline
        $m_{\rm pl,1}$ & $R_{\rm pl,1}$ & $m_{\rm pl,1}^{\rm Fe}$ & $R_{\rm pl,1}^{\rm Fe}$ & $f_{\rm Fe}$\\
        ($M_{\oplus}$) & (km) & ($M_{\oplus}$) & (km) & (\%) \\
        \hline
        9.8 & 11634.8 & 4.6 & 6507.8 & 47  \\
        20.7 & 13800.4 & 10.5 & 7744.1 & 51 \\
        \hline
    \end{tabular}
\end{table}

\subsection{Orbital parameters of planets}

The orbital parameters of the planets could be closely related to
their origin. First, orbit configuration with $0 \lesssim
e \lesssim 1$ and $a \gtrsim 1$ can form through the capture of a
planet by a WD-planet system \citep[e.g.,][]{Goulinski2018MNRAS}.
Second, the orbital parameters may be affected by the evolution
history of the WD's main sequence (MS) progenitor. For $M_{\rm
WD}\sim 0.6M_{\sun}$, the mass of its progenitor star should be
$M_{\rm MS}\sim 1.5 M_{\sun}$ according to
\citet{Cummings2018ApJ}. The envelope of such a star expands to
about 0.8--2 au during the red giant branch (RGB) and
asymptotic giant branch (AGB) stage \citep{Mustill2012ApJ, Villaver2014ApJ,
Veras2016MNRAS.463.2958V, Veras2020MNRAS.492.6059V}. Low-mass
planets initially resided inside the maximum stellar radius can
still survive since their orbits will expand due to the mass loss of
the progenitor during the giant branch phase
\citep{Mustill2012ApJ}. For a planet in a near-circular orbit, the
adiabatic stellar mass loss causes the orbit to expand by a
magnitude of approximately $M_{\rm MS}/M_{\rm WD}=2.5$ and finally
triggers dynamical instabilities, i.e. planetary systems that are
stable in the MS stage become unstable at the onset of the WD
stage \citep{Debes2002ApJ, Mustill2014MNRAS, Veras2016RSOS}.
During the unstable phase, ejection, collision, and scattering
processes may occur till the separations between the remaining
planets become wide enough so that the system be stabilized again
\citep{Adams2003Icar, Mustill2014MNRAS, Mustill2018MNRAS, 
	Frewen2014MNRAS, Veras2016MNRAS.458.3942V, Maldonado2020bMNRAS,
	Maldonado2021MNRAS, Maldonado2022MNRAS}.
A planet with an initial semi-major axis of
$\sim 2$ au at the MS stage may expand to an orbit of $\sim 5$ au
due to mass loss of the progenitor. The planet could migrate
inward to an orbit of $\sim 2$ au again due to recurrent close
scattering, which has been proved by numerical simulations for
planetary systems composed of equal-mass planets
\citep{Nagasawa2011ApJ}.

In short, orbital configurations with $0 \lesssim e
\lesssim 1$ and $a \gtrsim 1$ are possible in realistic cases
thanks to various dynamical interactions such as scattering and/or
capturing at the WD stage. In this study, we take two different
orbital configurations as: (A) $a_{1} = 3$ au, $a_{2} = 18$ au,
$e_{1} = 0.8$, $e_{2} = 0.3$; (B) $a_{1} = 6$ au, $a_{2} = 44.1$
au, $e_{1} = 0.8$, $e_{2} = 0.3$.

\subsection{Orbital evolution}

The dynamically stable configurations formed from
planet-planet scattering or capturing in a WD-planet system will
further evolve via the secular effect, the Kozai-Lidov mechanism.
In this secular evolution process, the semi-major axis remains
nearly constant while the eccentricity oscillates.
We now assess the stability of the planetary system. The stability
of triple systems has been studied for a long time
\citep{Eggleton1995ApJ,Mardling2001MNRAS}. The stability condition
can be expressed as \citep[e.g.,][]{He2018MNRAS}
\begin{align}\label{stablity}
\frac{a_{2}\left( 1 - e_{2} \right) }{a_{1}} > 2.8
\left[ \left(1+\frac{m_{\rm pl,2}}{M_{\rm WD} + m_{\rm pl,1}} \right) \frac{1+e_{2}}{(1-e_{2})^{1/2}}\right]^{2/5} \nonumber \\\times
\left( 1 - \frac{0.3i}{180^{\circ}}\right).
\end{align}

\begin{figure*}
    \plotone{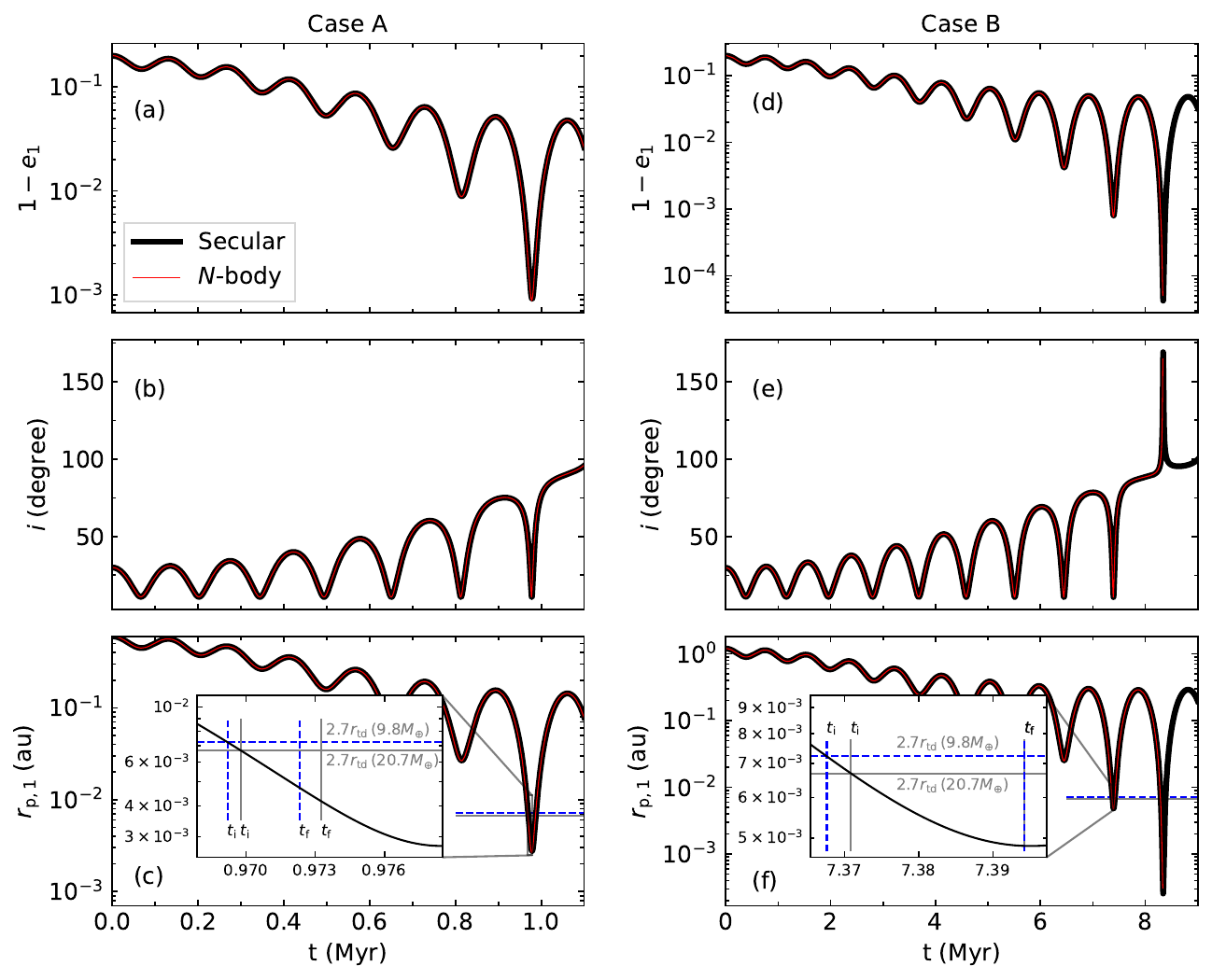}
    \caption{Evolution of the eccentricity (Panels a and d),
    inclination (Panels b and e), and periastron distance (Panels c and f)
    for the inner orbit planet (test particle) in a triple system.
    The mass of the central WD is taken as $M_{\rm WD} = 0.6 M_{\sun}$
    and the outer orbit planet has a mass of $m_{\rm pl,2} = 9.55\times10^{-4} M_{\sun} (1 M_{\rm Jup})$.
    The initial values of the argument, ascending node, and relative inclination of the two orbits
    are taken as
    $\omega_{1} = 180^{\circ}$, $\omega_{2} = 0^{\circ}$,
    $\Omega_{1} = 0^{\circ}$, $\Omega_{2} = 0^{\circ}$, and $i = 30^{\circ}$.
    Two cases are considered for the initial values of the semi-major axis and eccentricity.
    Case A (left panels): $a_{1} = 3$ au, $a_{2} = 18$ au, $e_{1} = 0.8$, $e_{2} = 0.3$.
    Case B (right panels): $a_{1} = 6$ au, $a_{2} = 44.1$ au, $e_{1} = 0.8$, $e_{2} = 0.3$.
    In each panel, the black and red lines show the results of the secular and \emph{N}-body simulations, respectively.
    But note that they are almost overlapped since the results are very close to each other.
    The dashed and solid horizontal lines represent the expected partial disruption distances
    for $m_{\rm pl,1} = 9.8 M_{\oplus}$ and $m_{\rm pl,1} = 20.7 M_{\oplus}$, respectively.
    The dashed and solid vertical lines denote the corresponding starting ($t_{\rm i}$)
    and stopping ($t_{\rm f}$) time of partial disruption. In Case B, $t_{\rm f}$ is essentially
    the same for different masses.}
\label{fig:fig3}
\end{figure*}

The parameters of the planetary systems considered in our
study (both Case A and Case B) satisfy this stability condition so
that the triple systems are stable on long timescales. In the
subsequent evolution, the orbital elements of the planet in the
inner orbit (with mass $m_{\rm pl,1}$) periodically change due to
the perturbation from the planet in the outer orbit (with mass
$m_{\rm pl,2}$) via the Kozai-Lidov mechanism. Since $m_{\rm pl,1}
\ll M_{\rm WD}$, the inner orbit planet can be regarded as a test
particle. To calculate the orbital evolution of the system, we use
the secular code\footnote{The code can be downloaded at
\url{https://github.com/bhareeshg/gda3bd}} developed by
\citet{Bhaskar2021AJ}, which adopts the double-averaging method
and can accurately describe the evolutionary properties of
hierarchical and mildly hierarchical systems. We take 0.01\% of
the Kozai timescale $t_{\rm k} = \frac{M_{\rm WD}}{m_{\rm
pl,2}}(1-e_{2}^{2})^{3/2}(\frac{a_{2}}{a_{1}})^{3}P_{\rm orb,1}$
as the time step in our secular calculations (i.e., $10^{-4}t_{\rm
k}$). The secular results are compared with the standard
\emph{N}-body simulation results acquired by using the package
$\mathrm{MERCURY}$ \citep{Chambers1999MNRAS}. The Bulirsch-Stoer
integration algorithm with a time step of 5\% of the period of the
inner orbit planet (i.e., 0.05$P_{\rm orb,1}$) and a tolerance
parameter of $10^{-12}$ are used for \emph{N}-body simulations. As
shown in Figure \ref{fig:fig3}, the secular results consist well
with the \emph{N}-body simulation results. It can be seen that
the periastron distance ($r_{\rm p,1} = a_{1}(1 - e_{1})$) of the
inner orbit planet decreases periodically to a small value so that
it could be partially disrupted at the periastron.

\section{Evolution of the inner orbit planet}\label{sec:mass-evolution}

The periodic change of the periastron distance leads to a continuous decrease of the
mass of the inner orbit planet since it would be partially disrupted every time it
passes through the periastron. The mass of the surviving portion can be evaluated
by considering the $r_{\rm p,1}$ parameter and the partial disruption criteria.

\subsection{Tidal disruption of a planet} \label{subsec:tidal-radius}

If a planet is located too close to its host WD, it will be tidally disrupted
since the host's tidal force is larger than the planet's self-gravity
at its surface \citep{Hills1975Natur,Rees1988Natur}. For a gravity-dominated
planet, the characteristic tidal disruption radius can be estimated as
\begin{eqnarray} \label{eq:r_td}
    r_{\rm td} = R\left( \frac{2M_{\rm WD}}{m}\right)^{1/3},
\end{eqnarray}
where $m$ and $ R $ are the mass and radius of the planet, respectively.
When the separation between the planet and the central WD ($r$)
is smaller than $r_{\rm td}$, it would be completely disrupted. But in the
more general cases that $r$ is slightly larger than $r_{\rm td}$, a partial
disruption will occur. It has been shown that the partial disruption occurs
as long as $ r < 2.7\,r_{\rm td} $ \citep{Guillochon2011ApJ, Liu2013ApJ}.
Such a process has also been explored through numerical
simulations \citep[e.g.,][]{Guillochon2013ApJ, Malamud2020a, Ryu2020ApJ, Law-Smith2020ApJ, Coughlin2022MNRAS}.
Astrophysical phenomena possibly connected with partial disruption have also
been reported \citep{Manser2019Sci}.
The properties of tidal interaction between a planet and a compact star
(neutron star or strange quark star) have also been investigated for different
scientific purposes such as searching for strange quark matter planets
\citep{Geng2015ApJ_a,Huang2017ApJ,Kuerban2019AIPC,Kuerban2020ApJ},
fast radio bursts \citep{Kurban2022ApJ} and repeating X-ray bursts \citep{Kurban2024AA}.

For the highly elliptical orbit considered in this study, $r$ is phase-dependent
and varies in a very wide range. The planet is affected by the tidal effect mainly
near the periastron, $ r_{\rm p,1} = a_{1}(1 - e_{1}) $, and it is relatively safe
at other orbital phases. When the partial disruption condition is satisfied, i.e.
$r_{\rm td} < r < 2.7\,r_{\rm td}$, the surface material of the planet will be
stripped off but the underneath portion will retain its integrity.
In our cases, the inner orbit planet will be subject to partial disruption
when its periastron distance decreases to $ r_{\rm p,1} = 2.7\,r_{\rm td} $ during
the orbit evolution process.
For example, the dashed and solid horizontal lines in
Figure \ref{fig:fig3} represent the partial disruption distances
for $m_{\rm pl,1} = 9.8 M_{\oplus}$ and $m_{\rm pl,1} = 20.7 M_{\oplus}$,
respectively. The dashed and solid vertical lines denote the corresponding
starting ($t_{\rm i}$) and stopping ($t_{\rm f}$) time of partial disruption.
It can be seen that for planets with different masses, the
partial disruption process stops at different $t_{\rm f}$ in Case A, while
it stops at the same $t_{\rm f}$ in Case B.

\subsection{Mass loss}\label{subsec:mass-loss}

During the partial disruption process, the total mass loss from the Lagrangian
points $ L_{1} $ and $ L_{2} $ in an encounter
is $ \Delta m = \Delta m_{\rm 1} + \Delta m_{\rm 2} $, where $ \Delta m_{\rm 1} $
and $ \Delta m_{\rm 2} $ are the mass loss from the $ L_{1} $ and $ L_{2} $,
respectively. Usually, the mass loss is asymmetric, and approximately 75\%
of $\Delta m$ is stripped from $L_{1}$ \citep{Faber2005Icar},
i.e. $ \Delta m_{\rm 1} \approx 0.75 \Delta m > \Delta m_{\rm 2} $. In our cases,
$ \Delta m$ depends on both the periastron separation and the planet's structure, which vary with time.
As mentioned earlier, we have $r_{\rm p,1} = r_{\rm td}/\beta$ for the partial
disruption condition, where $\beta = 1/2.7$. Then the mass evolution of
the surviving portion of the planet can be expressed as
\begin{equation}
    \frac{dm}{dt} = \frac{3r_{\rm p,1}^2\beta^3}{2M_{\rm WD}}\left(\frac{3R^2}{m}\frac{dR}{dm} - \frac{R^3}{m^2}\right)^{-1} \frac{dr_{\rm p,1}}{dt},
\end{equation}
where $dr_{\rm p,1}/dt$ is the varying rate of planet's periastron distance.

Taking the planet mass as $m_{\rm pl,1} = 9.8 M_{\oplus}$ or $20.7 M_{\oplus}$,
we have solved the above equation numerically to determine the mass loss of the
inner orbit planet for the cases of A and B. Here, we use $\Delta t$ to denote the period that the
planet's periastron distance $r_{\rm p,1}$ changes
from $r_{\rm p,1}(t_{\rm i}) $ to $r_{\rm p,1}(t_{\rm f})$.
It corresponds to the time interval between the starting ($t_{\rm i}$)
and stopping ($t_{\rm f}$) time of the partial disruption
process, $\Delta t = t_{\rm f} - t_{\rm i}$. The initial values of mass and
periastron are taken as $m_{0} = m(t_{\rm i}) = m_{\rm pl,1}$
and $ r_{\rm p,1}^{0}= r_{\rm p,1}(t_{\rm i}) = r_{\rm td}(m_{0})/\beta$.
Note that the material configuration inside the planet does not change dramatically
after a partial disruption due to the tensile strength and internal
viscosity \citep[e.g.,][]{Veras2019MNRAS.486.3831V}. The
total number of orbital periods within the time interval $\Delta t$
is $N = \Delta t/P_{\rm orb,1}$.

\begin{figure}
    \plotone{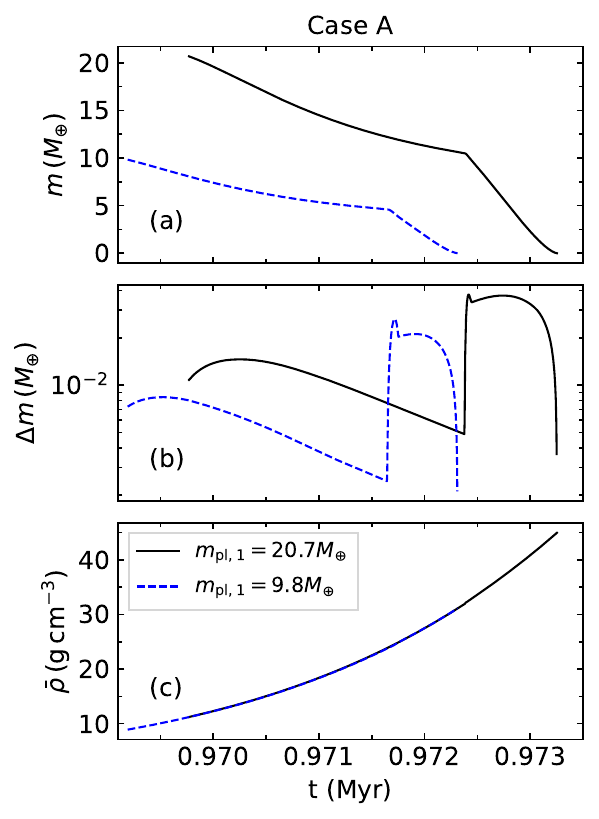}
    \caption{Evolution of the inner orbit planet during the partial disruption process for
    Case A: (a) Mass of the surviving portion, (b) mass loss, and (c) mean density.
    The dashed and solid lines correspond to the different initial masses of the planet,
    $9.8 M_{\oplus}$ and $20.7 M_{\oplus}$, respectively.}
    \label{fig:fig4}
\end{figure}

\begin{figure}
    \plotone{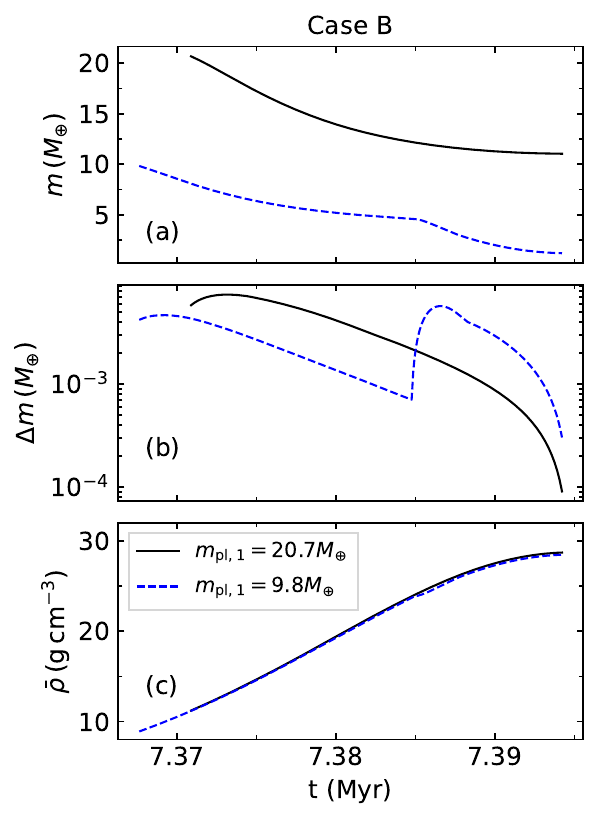}
    \caption{Same as in Figure \ref{fig:fig4} but for Case B.}
    \label{fig:fig5}
\end{figure}

Figures \ref{fig:fig4} and \ref{fig:fig5} show our numerical
results corresponding to cases A and B, respectively, which
illustrate the evolution of the planet structure due to the mass
loss during the close approach. It can be seen from the Figures
that the surviving portion of the planet becomes smaller and
smaller over a timescale of thousands of years, resulting in an
increase in the mean density. Consequently, the percentage of the
planet's iron composition increases at the final stages.
Note that there is a large difference in the mass loss
between Case A and Case B. This is mainly caused by the
different orbit evolution and different planetary structures.
The two planets in Case A will experience complete disruption at
different $t_{\rm f}$. However, the remnant core survives in
Case B and the partial disruption stops at the same $t_{\rm f}$
for the two objects. Note that the upward jumps in the mass
loss curves are caused by the change of $dR/dm$ at the
core-mantel boundary. The planet with $m_{\rm pl,1} = 20.7 M_{\oplus}$
in Case B keeps a fraction of its mantel so that its mass loss curve
is different from others.

\begin{table}[h]
    \caption{Key parameters derived for the two cases of orbit configurations.\label{tab:table2}}
    \centering
    \begin{tabular}{lccccc}
        \hline\hline
        Orbit conf. &$m(t_{\rm i})$ &$m(t_{\rm f})$ & $\Delta t$ & $N$ & $\Delta M $ \\
        & ($M_{\oplus}$) & ($M_{\oplus}$) & (yr) & (Orbit) & ($M_{\oplus}$) \\
        \hline
        Case A  &9.8  & 0    & 3122 & 465 & 9.8  \\
                &20.7 & 0    & 3481 & 519 & 20.7 \\
        Case B  &9.8  & 1.2  & 26600 & 1401   & 8.6  \\
                &20.7 & 11.1 & 23341  & 1229   & 9.6  \\
        \hline
    \end{tabular}
\end{table}

From these calculations, we see that partial disruptions could
occur repeatedly during the long-term evolution of the planet's
orbit. The total mass loss ($\Delta M$), which is defined as the
change of the planet mass during the time interval $\Delta t$, can
be calculated as $ \Delta M = m(t_{\rm i}) - m(t_{\rm f})$, where
$m(t_{\rm i}) = m_{\rm pl,1}$ is the initial mass and $m(t_{\rm
f})$ is the final mass after $\Delta t$. Table \ref{tab:table2}
presents our numerical results for these parameters.
In Case A, we have $\Delta t$ = 3122 yr ($N = 465 $) and $ \Delta M =
9.8 M_{\oplus}$ for $m_{\rm pl,1} = 9.8 M_{\oplus}$, while $\Delta
t$ = 3481 yr ($N = 519 $) and $ \Delta M = 20.7 M_{\oplus}$ for
$m_{\rm pl,1} = 20.7 M_{\oplus}$. In Case B, we have $\Delta t$ =
26600 yr ($N = 1401 $) and $ \Delta M = 8.6 M_{\oplus}$ for
$m_{\rm pl,1} = 9.8 M_{\oplus}$, while $\Delta t$ = 23341 yr ($N =
1229 $) and $ \Delta M = 9.6 M_{\oplus}$ for $m_{\rm pl,1} = 20.7
M_{\oplus}$. It shows that: (i) the planet will become an
iron-rich or pure iron object via mass loss, and (ii) the mass
loss process depends on the structure of the planet and the
orbital configuration of the planetary system.


\section{Accretion of clumps}\label{sec:accretion}

Numerical simulations on the tidal disruption of a planet around a WD
\citep{Malamud2020a,Malamud2020b} show that the size of the generated
clumps range from a few kilometers to hundreds of kilometers.
The fate of the clumps produced during a partial disruption is different
from that produced in a full disruption. A full disruption usually occurs
in the cases of deep encounters which would lead to the formation of a ring/disc
so that the generated clumps evolve mainly under the influence of non-gravitational
effects \citep{Veras2014MNRAS,Veras2015MNRAS.451.3453V,Hogg2021MNRAS,Zhang2021ApJ}.
However, the clumps produced during a partial disruption process are affected by the gravitational
perturbation from the remnant planet. They could be further tidally disrupted
when they move closer to the WD. Note that these small bodies are
generally dominated by intrinsic material strength but not gravity. As a
result, the breakup separation is $\sim 10^{9}$ cm for
them \citep[e.g.,][]{Zhang2021ApJ,Kurban2023MNRAS}. In this section, we
will investigate the evolutionary tracks of these clumps.

\subsection{Orbital parameters of the clumps}\label{subsec:para}

In our framework, the orbit of the inner planet evolves with time. Here, we analyze the
distribution of the orbits of the clumps generated during the partial disruption process.
The semi-major axis of the clumps can be calculated by following \citet{Malamud2020a},
\begin{equation}\label{a_prime}
    a_{\rm cl} = \left\{
    \begin{array}{lr}
        a_{1} \left( 1 + a_{1}\frac{2R}{d(d - R)}\right)^{-1},~~(\rm inner)\\
        a_{1} \left( 1 - a_{1}\frac{2R}{d(d + R)}\right)^{-1},~~(\rm outer)
    \end{array}
    \right.
\end{equation}
where $ a_{1} $ is still the inner orbit planet's original semi-major axis, $ d $ is
the distance between the WD and the planet at the moment of break
up (i.e., $d = r_{\rm p,1} = r_{\rm td}(m)/\beta$, with $m$ being the mass of
the remnant planet that evolves with time), $ R $ is the displacement of the clump
relative to the planet's mass center at the moment of breakup. Actually, $R$ is
the radius of the remnant planet here ($ R = 0 $ corresponds to the planet's center).
In the above equation, ``inner'' means the clumps originate from the inner side of
the planet, while ``outer'' means they come from the outer side. The clumps in the
inner stream are all bound to the WD. For clumps originating from the
planet's face opposite to the WD, there is a critical displacement of
$ R_{\rm crit} = d^{2}/(2a_{1} - d) $. The clumps with $ R > R_{\rm crit} $ are
unbound while particles with $ R < R_{\rm crit} $ are still bound to the WD
\citep{Malamud2020a}. We have checked the properties of the clumps in
the outer stream (see Figure \ref{fig:fig1}) and found that their orbits are
bound to the WD when they are newly born from the remnant planet.
But their orbits could be altered by the planet after a few orbits and
may be scattered away finally. We will discuss this point in the next sections.

Combining Equation (\ref{a_prime}) and Kepler's third law, we can obtain
the orbital period of the clumps as
\begin{equation}
    P_{\rm orb}^{\rm cl} = \left( \frac{4 a_{\rm cl}^3 \pi^2 }{G (M_{\rm WD} + m_{\rm cl})}\right)^{1/2}.
\end{equation}
After the disruption, the clumps return to the periastron ($r_{\rm pl,1}$) at
a time of $ t_{\rm re} = P_{\rm orb}^{\rm cl} $.
Since $ r_{\rm p,1} = a_{\rm cl}(1 - e_{\rm cl}) \pm R $, the eccentricity of the
clumps can be expressed as
\begin{equation}\label{e_cl}
    e_{\rm cl} = \left\{
    \begin{array}{lr}
        1 - \frac{r_{\rm pl,1} - R}{a_{\rm cl}},~~(\rm inner)\\
        1 - \frac{r_{\rm pl,1} + R}{a_{\rm cl}},~~(\rm outer)
    \end{array}
    \right.
\end{equation}

\subsection{Fate of the clumps}\label{subsec:fate_of_clumps}

Clumps in the outer orbit are affected by the perturbation from both the remnant
planet and the outer orbit planet (whose mass is $m_{\rm pl,2}$). Their evolution
is very complicated. They could be ejected from the system or could have
chaotic behavior. Generally, it is hard to describe their properties in a simple
way. Here, we mainly concentrate on the inner orbit clumps in this study.

After the birth of the clumps, the WD, clumps, and the surviving portion
of the planet form a multi-body system, in which the gravitational interactions
among various objects are very complicated. For simplicity, we omit the interactions
between the clumps. To follow the evolution of a particular clump, we can take the clump,
the WD, and the surviving part of the planet as a triple system. The stability
of such a triple system can be accessed by using Equation (\ref{stablity}). It is found
that the configuration does not satisfy the stability condition. In fact, we
have $ a_{1}(1 - e_{1})/a_{\rm cl} < 0.2 $, which is well below the stability limit.
It means that the secular interaction (Kozai-Lidov mechanism) will be broken
down, i.e., Kozai-Lidov mechanism can not accurately describe clump's orbit evolution
\citep[e.g.,][]{Perets2012ApJ, Naoz2011Natur, Katz2012, Antonini2012ApJ, Bode2014MNRAS}.
In such a case, strong perturbation from the remnant planet \citep[e.g.,][]{Toonen2022AA} will
lead to a large orbital eccentricity for the clump so that it will lose its angular momentum
on a relatively short timescale \citep[e.g.,][]{Antonini2014ApJ, Antonini2016ApJ, Hamers2022ApJ, Kurban2023MNRAS}.
Below, we present a detailed estimation on this point.

The clump returns to the periastron in a time of $t_{\rm re} = P_{\rm orb}^{\rm cl}$
after the disruption \citep{Malamud2020a,Zanazzi2020MNRAS,Rossi2021SSRv}. Its specific angular momentum
is $j_{\rm cl} =\sqrt{1 - e_{\rm cl}^2}$, which is small compared with that
of the planet. Perturbation from the remnant planet leads to a quick loss of
the clump's angular momentum. The angular momentum changes from $j_{\rm cl}$
to almost zero on a timescale of the order of the inner orbit period, causing
the clump to effectively fall onto the WD \citep[e.g.,][]{Antonini2014ApJ, Kurban2023MNRAS}.
It indicates that the periastron of the clump jumps to a very small value, which is
mainly caused by the energy exchange between the clump and the remnant planet due to
their gravitational interactions \citep[e.g.,][]{Zhang2023ApJ}. The evolution
timescale of the angular momentum of the clump ($t_{\rm evo}$) can be expressed
as \citep[e.g.,][]{Antonini2014ApJ}
\begin{equation}\label{}
    \left[ \frac{1}{j_{\rm cl}}\frac{dj_{\rm cl}}{dt}\right]^{-1} \approx \frac{P_{\rm orb}^{\rm cl}}{5\pi}
    \frac{M_{\rm WD}}{m}\left[ \frac{a_{1}(1 - e_{1})}{a_{\rm cl}} \right]^3 \sqrt{1 - e_{\rm cl}}.
\end{equation}

Adopting the parameters used in Figures \ref{fig:fig4} and \ref{fig:fig5}, we have
estimated the orbital periods and the evolution timescales of the clumps' angular
momentum. Table \ref{tab:table3} lists the results at the time of $t_{\rm i}$ and
$t_{\rm c}$ in Case A, where $t_{\rm c}$ is the critical time at which
$m \approx 0.5 M_{\oplus}$ and $t_{\rm evo} < P_{\rm orb}^{\rm cl} < P_{\rm orb} $
for $t \lesssim t_{\rm c}$. For Case B, the results at the time of $t_{\rm i}$ and $t_{\rm f}$
are listed in Table \ref{tab:table4}. It can be seen that
$t_{\rm evo} < P_{\rm orb}^{\rm cl} < P_{\rm orb}$ for $t \lesssim t_{\rm f}$.
Since the clumps lose their angular momentum in a time
of $t_{\rm evo}$ after the second orbit (i.e., the second periastron passage),
the clump's total travel time from their birth at the planet to the WD
will be $t_{\rm trav} \lesssim 3 P_{\rm orb}^{\rm cl}$ \citep{Kurban2023MNRAS}.
It means that the clump could fall onto the WD in a short time.

\begin{table*}[t]
    \centering
    \caption{Case A: the orbital period and angular momentum loss time for the clumps generated
             at time $t_{\rm i}$ and $t_{\rm c}$, and the characteristic accretion rate during the time interval $\Delta t = t_{\rm c} - t_{\rm i}$.\label{tab:table3}}
    \begin{tabular}{cccc|cccc|c}
        \hline\hline
        $m(t_{\rm i})$ &$R(t_{\rm i})$ & $ P_{\rm orb}^{\rm cl}$ & $t_{\rm evo}$ & $m(t_{\rm c})$ &$R(t_{\rm c})$ & $ P_{\rm orb}^{\rm cl}$ & $t_{\rm evo}$& $\langle \dot{M} \rangle$ \\
        ($M_{\oplus}$) & (km) & (day) & (day) & ($M_{\oplus}$) & (km) & (day) & (day) &(g s$^{-1}$)\\
        \hline
         9.8  &   11634.8  &   76.9  &   0.2  &   0.5  &   2894.4  &   167.2  &   0.4  & 4.4e+17 \\
         20.7  &   13800.4  &   49.1  &   0.1  &   0.5  &   2558.4  &   141.6  &   0.3  & 8.5e+17 \\
        \hline
    \end{tabular}
\end{table*}

\begin{table*}[t]
    \centering
    \caption{Case B: the orbital period and angular momentum loss time for the clumps generated
        at time $t_{\rm i}$ and $t_{\rm f}$, and the characteristic accretion rate during the time interval $\Delta t = t_{\rm f} - t_{\rm i}$.\label{tab:table4}}
    \begin{tabular}{cccc|cccc|c}
        \hline\hline
        $m(t_{\rm i})$ &$R(t_{\rm i})$ & $ P_{\rm orb}^{\rm cl}$ & $t_{\rm evo}$ & $m(t_{\rm f})$ &$R(t_{\rm f})$ & $ P_{\rm orb}^{\rm cl}$ & $t_{\rm evo}$& $\langle \dot{M} \rangle$ \\
        ($M_{\oplus}$) & (km) & (day) & (day) & ($M_{\oplus}$) & (km) & (day) & (day) &(g s$^{-1}$)\\
        \hline
        9.8  &   11634.8  &   83.1  &   0.2  &   1.2  &   3909.8  &   130.4  &   0.2   & 4.6e+16 \\
        20.7  &   13800.4  &   51.9  &   0.1  &   11.1  &   8183.3  &   44.8  &   0.1 & 5.9e+16 \\
        \hline
    \end{tabular}
\end{table*}

\subsection{Accretion rate} \label{accretion-rate}

\begin{figure*}[t]
    \plotone{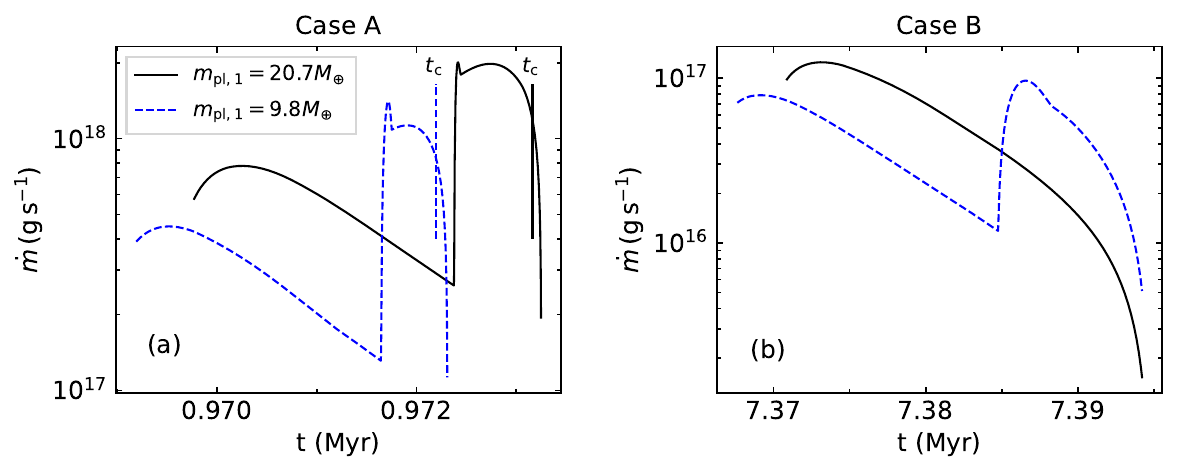}
    \caption{Accretion rate as a function of time for Case A (Panel a) and
     Case B (Panel b). In each panel, the dashed and solid line
     correspond to $m_{\rm pl,1} = 9.8 M_{\oplus}$
     and $m_{\rm pl,1} = 20.7 M_{\oplus}$, respectively. In Panel (a),
     the vertical lines marked with $t_{\rm c}$ represent the critical
     time at which $m \approx 0.5 M_{\oplus}$.}
    \label{fig:fig6}
\end{figure*}

The clumps would fall onto the WD on a timescale shorter than a
few $ P_{\rm orb}$ after their birth. It means that the lost mass
of the planet during an encounter, $\Delta m_{1}$, would be
accreted by the WD in this period. We thus can simply estimate the
accretion rate. Figure \ref{fig:fig6} shows the long-term
evolution of the accretion rate ($\dot{m} = 0.75dm/dt$) for our
cases A and B. We see that the accretion rate is
$\dot{m} \sim10^{17}$--$10^{18}$ g s$^{-1}$ in Case A and
$\dot{m} \sim10^{15}$--$10^{17}$ g s$^{-1}$ in Case B. It
depends on the planet's structure and the periastron distance of the
orbit. In both cases, the accretion rate changes in a complex way
because $dr_{\rm p,1}/dt$ and $dR/dm$ vary in the process.

The total mass loss is $\Delta M = \Delta M_{\rm 1} + \Delta
M_{\rm 2}$, where $ \Delta M_{1} $ is the sum of the mass lost
from $L_{1}$ and $ \Delta M_{2} $ is the sum of the masses lost
from $ L_{2} $, respectively. The average accretion rate can be
simply calculated as $\langle \dot{M} \rangle = \Delta M_{\rm 1} /
\Delta t \approx 0.75 \Delta M/ \Delta t $ (here $\Delta t =
t_{\rm c} - t_{\rm i}$ and $\Delta M = m(t_{\rm i}) - m(t_{\rm
c})$ for Case A). The characteristic accretion rate for cases A
and B are presented in Table \ref{tab:table3} and
\ref{tab:table4}, respectively. In Case A, $\langle
\dot{M} \rangle \approx 4.4\times10^{17}$ g s$^{-1}$ for $m_{\rm
pl,1} = 9.8 M_{\oplus}$, and $\langle \dot{M} \rangle \approx
8.5\times10^{17}$ g s$^{-1}$ for $m_{\rm pl,1} = 20.7 M_{\oplus}$.
In Case B, $ \langle \dot{M} \rangle  \approx 4.6\times10^{16}$ g
s$^{-1}$ for $m_{\rm pl,1} = 9.8 M_{\oplus}$, and $ \langle
\dot{M} \rangle  \approx 5.9\times10^{16}$ g s$^{-1}$ for $m_{\rm
pl,1} = 20.7 M_{\oplus}$. When the planet mass is equal, the
accretion rate in Case A is generally higher than that in Case B.
The reason is that the decrease of $r_{\rm p,1}$ in Case A is
quicker than that in Case B during the partial disruption phases.
The results show that the accretion rate is also governed by the
planet's structure and orbital configuration, which is similar to
the mass loss.

\section{Discussion}\label{sec:discussion}

Depending on the planetary architecture, there are three
possibilities for the final fate of the planet subjected to a
long-term partial disruption process due to the orbital evolution
caused by the Kozai-Lidov mechanism. Firstly, the planet could be
destroyed completely before $t_{\rm f}$, which may lead to the
formation of a relatively massive remnant disc. Secondly, the main
portion of the planet survives after a large number of partial
disruptions during $ \Delta t$. After $t_{\rm f}$ it could be
further disrupted if $r_{\rm p,1}$ still meets the partial
disruption condition during the next high eccentricity
phase of the Kozai-Lidov cycle, as in Case B. 
Thirdly, for some particular configurations, the remnant
planet survives at $t_{\rm f}$ and acquires its maximum eccentricity
($e_{\rm max}$) at this point. In subsequent evolution, the eccentricity
$e$ remains less than $e_{\rm max}$. Consequently, the partial disruption
distance of the remnant planet will always be less than its periastron
distance for $t > t_{\rm f}$ so that the tidal disruption process will
no longer occur later. Our model extends the timescale across which a
planet can be disrupted compared to a complete instantaneous tidal disruption
at the Roche radius. Here we present some further analyses of the
process.

\subsection{Iron-rich planetary remnant}

In the multiple partial disruption process, the planet's mass gradually decreases
together with the decrease of the periastron distance. The surface layer of the
planet is stripped off at each periastron passage so that the percentage of iron
composition in the remnant planet becomes higher and higher. The iron core may be
surrounded by a layer of debris at the late stages of the partial disruption process.
Such iron-rich objects have been discovered in recent years. For example,
the minor planet SDSS J1228+1040 b was found to be embedded in a debris disc. The planet
is about 0.003 au from its host star, well within the Roche radius of rubble-pile.
However, it does not show any signatures of being disrupted now, implying that the
object is actually an iron-rich planetary core with tensile strength and internal
viscosity \citep{Veras2019MNRAS.488..153V, Manser2019Sci, Connor2020MNRAS}.

Iron-rich planetary objects have also been discovered around other type of stars.
For example, the terrestrial exoplanet GJ 367 b is found to be orbiting around
an M-type star \citep{Lam2021Sci, Goffo2023ApJ}.
It has a mass of $0.633 M_{\oplus}$, a radius of $0.699 R_{\oplus}$,
and a mean density of $\bar\rho$ = 10.22 g cm$^{-3}$, implying that its iron
composition is about 90\%. Its distance from the host star is 0.00709 au, which
is well within the rubble-pile Roche radius. Interestingly,
GJ 367 has two additional planets, GJ 367 c ($> 4.13 M_{\oplus}$)
and GJ 367 d ($> 6.03 M_{\oplus}$), residing outside the GJ 367 b \citep{Goffo2023ApJ}.
Another exoplanet K2-141 b, which orbits around a K-type star, has a mass
of $4.97 M_{\oplus}$, a radius of $1.51 R_{\oplus}$, and a mean density
of $\bar\rho$ = 7.96 g cm$^{-3}$.  It is 0.00747 au from its host star,
which is also well within the rubble-pile Roche radius \citep{Malavolta2018AJ}.
K2-141 has another additional planet, K2-141 c (mass: $7.41 M_{\oplus}$), which also
resides outside K2-141 b \citep{Malavolta2018AJ}. The formation history of the
iron-rich inner orbit planets in the above two planetary systems (GJ 367 and K2-141)
is still not clear at present. A likely interpretation is that they may have
experienced the same evolutionary history as in our framework.

\subsection{Total accreted mass}

The masses of the bodies that pollute WDs can be
inferred from observations. For example, accumulated metal mass in
the convective zone can be estimated by using the spectral
features and WD atmospheric models \citep{Koester2009AA}.
\cite{Girven2012ApJ} derived the lower limit of the total accreted
mass of metals deposited in the convective zone as $\sim 10^{16}$--
$10^{22}$ kg for DBZ WDs. \cite{Harrison2021MNRAS} also argued
that DZ WDs systems in which the accretion process has come to an
end can be unique probes for estimating the mass of the polluting
body. They found that the mass of the pollutant body is generally
consistent with that of large asteroids and/or the Moon.

However, since the tidal disruption, accretion, and
sinking of accreted materials is a complicated process, the mass
of metals in the convective zone is not necessarily equivalent to
the mass of the polluting body in some cases. Firstly, a fraction
of the planet's mass may escape from the system after the tidal
disruption. In the complete tidal disruption cases, some materials
could be unbounded and would be ejected \citep{Malamud2020a,
	Malamud2020b}. In the partial disruption cases, materials in the
outer bound stream may also be scattered away from the system due
to perturbation from the remnant planet \citep{Kurban2023MNRAS,
	Kurban2024AA}. Secondly, some of the energy released during the
accretion process will be thermalized and emitted in the form of
X-rays \citep{Cunningham2022Natur, Estrada-Dorado2023ApJ}, which
may more or less lead to an extra mass loss. Thirdly, some of the
accreted mass may sink too quickly and be deposited much deeper
than the upper convective zone so that they are no longer
observable. In fact, metals heavier than carbon will pass through
the convection zone quickly, on a timescale ranging from days to
weeks for DA WDs, or in a few Myr for DB WDs
\citep{Wyatt2014MNRAS}. This will cause an underestimation of the
total accreted mass based on observations involving only the
convective zone. Finally, some of the mass may be left in a
remnant accretion disc. For example, the accretion lasts for a
long period in most of DA WDs \citep{Kleinman2013ApJS,
	Veras2016RSOS, Hollands2018MNRAS, Blouin2022MNRAS}, implying that
large planets should be involved in the disruption process
\citep{Hamers2016MNRAS, Petrovich2017ApJ}.

\subsection{Accretion rate confronted with observations}

The accretion rate inferred from the observed photospheric
element abundances of DAZ WDs typically ranges from $\sim 10^5$ g
s$^{-1}$ to $\sim 10^8$ g s$^{-1}$ \citep{Koester2014AA}. A high
rate up to $\sim10^{11}$ g s$^{-1}$ is observationally inferred
for a number of DBZ WDs \citep{Farihi2012MNRAS, Girven2012ApJ,
Farihi2016NewAR}. The accretion rate derived from the X-ray
observations of WD G 29-38 is approximately $2 \times 10^9$ g
s$^{-1}$ \citep[\emph{Chandra}
observation;][]{Cunningham2022Natur} or $4.01 \times 10^9$ g
s$^{-1}$ \citep[\emph{XMM-Newton}
observation;][]{Estrada-Dorado2023ApJ}. An instantaneous high
accretion rate of $1.45 \times 10^{14}$ g s$^{-1}$ was also
derived from the X-ray luminosity of KDP 0005+5106 \citep{Chu2021ApJ}. 
When energy supply is assumed to come only from accretion,
heating of the Ca \textsc{ii} emission line region requires an
accretion rate of $10^{17}$--$10^{18}$ g s$^{-1}$ \citep{Hartmann2011AA}.
Note that this accretion rate can be lowered when considering the
heating by energy dissipation through disc asymmetries or by
absorbing photons from the WD \citep{Hartmann2011AA}.

Several models have been developed to explain the above
observations. The Poynting-Robertson drag and/or the Yarkovsky
effect can account for an accretion rate up to $\sim 10^{8}$ g
s$^{-1}$ \citep{Rafikov2011ApJ}. Considering the additional
viscosity of gas produced from sublimation would further increase
the accretion rate, potentially leading to a runaway process with
a peak rate of $10^{10}$--$10^{11}$ g s$^{-1}$
\citep{Rafikov2011MNRAS, Metzger2012MNRAS}. On the other hand, if
the collisional grind-down is taken into account, the peak
accretion rate is found to vary from $\sim 10^{8}$ g s$^{-1}$ for
a 50 km asteroid to $\sim 10^{13}$ g s$^{-1}$ for a 500 km
asteroid \citep{Brouwers2022MNRAS}. It was also argued that an
accretion rate of $\sim 10^{10}$ g s$^{-1}$ could be attained due
to the Alfv\'{e}n-wave drag \citep{Zhang2021ApJ}. The high
accretion rate associated with star-planet interaction can lead to
bright X-ray emissions \citep{Farihi2012MNRAS, Girven2012ApJ}. The
accretion rate of $1.45 \times 10^{14}$ g s$^{-1}$ inferred in KDP
0005+5106 strongly indicates the interaction of the WD with a
Jupiter-like planet \citep{Chu2021ApJ}, which is still two orders
of magnitude smaller than the lower limit of WDs accreting from
stellar companions at a rate of $\gtrsim10^{16}$ g s$^{-1}$
\citep{Mukai2017PASP}. As delineated in Section \ref{accretion-rate},
partial disruption of a planet can engender such high accretion rates.
The accretion rate of $10^{17}$--$10^{18}$ g s$^{-1}$ inferred from
the Ca \textsc{ii} emission lines is likely an overestimate \citep{Hartmann2011AA}
and it is the only value in the literature close to the accretion
rate found from our analysis. In the case of an instantaneous full
disruption, the accretion rate will be even much higher since
sufficient fragments will be generated in mutual collisions
\citep[e.g.,][]{Li2021MNRAS, Brouwers2022MNRAS}. Details of such
an extensive accretion process, which is out of the scope of this
study, still need to be clarified in the future.

Most of the observed accretion rates in polluted WD systems
are significantly lower than the rate in our work. This can be
attributed to the rarity of the relevant process.
The percentage of polluted WD systems, which is about 25\%--50\%
\citep{Zuckerman2003ApJ, Koester2014AA}, is roughly consistent with
the estimation of the exoplanet systems in the Milky Way \citep{Cassan2012Natur}.
The accretion rate decreases slowly
\citep{Hollands2018MNRAS, Chen2019NatAs} or remains relatively
steady \citep{Blouin2022MNRAS} throughout the whole cooling stage of WDs,
implying that the materials are continuously supplied.
This is hard to explain by a single asteroid disruption
event but requires extended debris from the tidal disruption
or interaction with an asteroid belt.
The planets around WDs can be involved in these processes, i.e.,
they can act either as pollutants or as drivers of pollutants depending
on their orbits. As mentioned previously, asteroid disruption models, which require
an object at least as massive as that of Earth's moon for pollution
drivers \citep{Veras2023MNRAS}, can produce low accretion rates.
When a planet plays the role of pollutant, the mass of the driver
(perturber) should be equal or larger than that of the pollutant to deliver it
to the tidal disruption radius. The absence of observations that are
in line with the accretion rates for the partial tidal disruption of a planet
suggests that the proposed mechanism might be a rare process. This is consistent
with the argument that the pollution resulting from the tidal disruption of
a planet is rare compared to the pollution from small bodies such as
asteroids \citep{Veras2024RvMG}.

Moreover, it is worth noting that the interaction between the
accreted materials and the WD may produce X-ray emission. The
gravitational potential energy of the accreted material can be
transformed into X-rays in the process. The corresponding X-ray
luminosity can be expressed as \citep{Cunningham2022Natur}
\begin{equation}\label{L_X}
	L_{\rm X} = A\frac{1}{2}\frac{GM_{\rm WD}\dot{m}}{R_{\rm WD}},
\end{equation}
where $A = 0.5^{+0.3}_{-0.4}$ is a constant relevant to plasma
cooling. It is usually assumed that half of the energy will be
absorbed by the star and another half will be emitted in X-rays,
which leads to the factor of 1/2 in the expression. Taken $M_{\rm
WD} = 0.6 M_{\odot}$, $R_{\rm WD} = 0.0129 R_{\odot}$, $A = 0.5$,
and $\dot{m} = 10^{14}$--$10^{18}$ g s$^{-1}$, we have an X-ray
luminosity of $L_{\rm X} = 2.2\times10^{30}$--$2.2\times10^{34}$ erg s$^{-1}$.
These X-ray luminosities are exceptionally high and exceed the expected
luminosity for a WD accreting material from an MS
companion \citep[$\sim10^{29}$--$10^{33}$ erg s$^{-1}$,][]{Cunningham2022Natur}.
This would be easier to observe than the previous X-ray observations of the
WD accretion process. The lack of previous detections of these phenomena
further implies that the proposed process is rare.

\subsection{Other forces that affect the clump evolution} \label{other-forces}

Other nongravitational effects such as the Poynting-Robertson drag, the Yarkovsky
effect, the Alfv\'{e}n-wave drag, and sublimation can affect the orbital
evolution of the tidal debris.
A study by \cite{Veras2014MNRAS} shows that a ring of debris (centimeter-sized)
can be formed during the tidal disruption of a rubble-pile asteroid.
The effect of Poynting-Robertson drag on the micrometer-to-cm-sized particles
could shrink and circularize the ring/disc effectively on a timescale ranging
from years up to hundreds of millions of years, depending on the exact size
of the debris and the age of the WD \citep{Veras2015MNRAS.451.3453V}.
The collisional grind-down in the tidal disc \citep{Wyatt2011CeMDA} can
facilitate the accretion of debris materials onto the WD due to
the Poynting-Robertson drag force \citep{Swan2021MNRAS,Li2021MNRAS,Brouwers2022MNRAS}.
However, when the debris particles are much larger than a centimeter size, the
Poynting-Robertson drag becomes ineffective, and instead the Yarkovsky effect
dominates \citep{Veras2015MNRAS.451.3453V,Veras2022MNRAS}.
Generally, the accretion timescale of small sub-cm or sub-dm particles
under the action of the radiative forces is as long as $ > 10^4 $ yr.

It has also been argued that a pre-existing compact gaseous disc in the
vicinity of a WD can facilitate the circularization of the orbit
of the tidal debris \citep{Malamud2021MNRAS, Brouwers2022MNRAS} and small
exo-Kuiper or exo-Oort like objects with a size
of 0.1--10 km \citep{Grishin2019MNRAS}. The circularization timescale
depends on the structure (mass and compactness) of the disc. In our study,
we have assumed that the circumstellar environment is clean. If a massive
gaseous disc exists before the planet is destroyed, the accretion of
the clumps will be accelerated correspondingly.

In addition to the above evolutionary routes, other effects may also play a
role in the long-term evolutionary processes. They include
the direct capture of asteroids/fragments by the WD due to the scatter
effect of the outer planet, the sublimation of debris material which occurs when
the debris is close to the star, and the accretion of gas
\citep{Veras2015MNRAS.452.1945V, Brouwers2022MNRAS}, and the gravitational
perturbation of outer planet \citep{Li2021MNRAS}. The timescales of these
effects are usually larger than $10^4$ yr. For example, the fragments are accreted
by the WD over several Myr under the influence of gravitational perturbation
of a distant planet with a semi-major axis of 10 au \citep{Li2021MNRAS}.

Alfv\'{e}n-wave drag caused by the magnetic field of the WD can
effectively circularize the orbit of circum-stellar debris discs \citep{Zhang2021ApJ}.
Usually, the circularization timescale depends on the size of the debris.
For larger particles, the timescale is longer than that of smaller ones.
For example, a fragment of 1 cm would be circularized on a timescale
of 10 yr, while a fragment of 10 km can be circularized on a timescale
of $10^{7}$ yr \citep{Zhang2021ApJ}.

In our framework, the fragments generated during the partial disruption
range from tiny dust to macroscopic asteroids. As mentioned previously,
the dynamics of micrometer-to-centimeter particles could be affected by
radiative force due to the Poynting-Robertson drag and Yarkovsky
effect \citep{Veras2015MNRAS.451.3453V}.
However, it is hard to estimate the exact mass fraction of these small
particles. On the other hand, according to \citet{Malamud2020a}, there
are at least two factors that prevent the formation of large amounts of
dusty materials. First, small particles can coagulate into much larger
fragments when they are gravitationally self-confined, which should be
orders of magnitude larger than the sizes relevant to either
the Poynting-Robertson drag or the Yarkovsky effect. Second, the particles lie
well within the gravitational potential of their parent fragments
(which are much larger) even if they are produced during the partial
disruption process. Thus the dominant force is still gravitational for
them, but not the Poynting-Robertson drag. Dusty particles could be
formed via the collisional grinddown in a tidal disc \citep{Wyatt2011CeMDA},
but the collisions are unlikely to occur at the early stage of the disruption.
Since the older fragments lose their angular momentum on a timescale of
several orbital periods due to perturbation from the remnant planet,
it is hard to form a disc before the complete destruction of the planet.
Therefore, the fragments are essentially largely unaffected by radiation effects.

The perturbation of the outer planet and other nongravitational effects such
as the Poynting-Robertson drag, the Yarkovsky effect, the Alfv\'{e}n-wave drag,
and sublimation may play a significant role when the inner planet is completely
destroyed. Collisional cascade in the tidal disc formed after the complete
destruction can produce a larger number of small dusty particles on a long
evolutionary timescale \citep{Wyatt2011CeMDA}. As a result, nongravitational
effects such as the Poynting-Robertson drag, the Yarkovsky effect, and the
Alfv\'{e}n-wave drag could be significant. The perturbation of the outer
planet on the tidal debris disc may also be long-term
effect \citep{Li2021MNRAS}. To summarize, the accretion is likely to proceed
under the combined effects of these channels. Each mechanism may play a
different role at different stages according to the detailed condition of
the system. But generally, at the early stages of the partial disruption, the
contribution of these nongravitational effects on the accretion is small
as compared with the gravitational perturbation from the remnant planet itself.


\section{Conclusions} \label{sec:conclusion}

In this study, we investigate the impact of gravitational perturbation on
the accretion of tidal debris in the case of partial tidal disruption. A
triple system composed of a WD, a rocky planet in the inner orbit,
and a Jupiter-like planet in the outer orbit is considered. The orbital
parameters of the system evolve under the influence of secular effects
such as the Kozai-Lidov mechanism. The eccentricity of the inner orbit
planet can be excited to extremely high values so that its periastron
distance decreases correspondingly, resulting in a partial disruption of
the inner orbit planet. The surface material of the planet is stripped
off but the main body can still survive. The partial disruption can
occur repeatedly over a few thousands of years. The surviving portion
of the planet loses its mass every time it passes through the periastron,
the rate of which depends on the initial orbital configuration of the
system and the planet's structure. A consequence of the mass loss is that
the percentage of iron composition and mean density of the remnant planet
increases, which leads the planet's core to finally become an iron-rich or
pure iron object.

For the clumps generated during a partial disruption, gravitational
perturbation from the remnant planet is a dominant force in the evolution.
The clumps lose their angular momentum on a short timescale and finally collide
with the WD. Again, the process is governed by the orbital
configuration of the system and the planet's structure. Other nongravitational
effects (the Poynting-Robertson effect, the Yarkovsky effect, the Alfv\'{e}n-wave
drag, and sublimation) can be neglected at early stages. However, gravitational
perturbation becomes weaker and weaker with the decrease of the remnant planet
mass. It may finally diminish if the remnant planet is destroyed completely.
The collisional grind-down in the tidal disc formed later in the disruption
can facilitate dust formation. It makes the nongravitational effects
significant at late stages.

The proposed mechanism in the work can produce a high accretion rate in
the range of $\sim 10^{14}$--$10^{18}$ g s$^{-1}$. The lack of observations of
such intense accretion episodes implies that the proposed mechanism in the work
might be a rare process. If such intense accretion events could be observed by
X-ray telescopes in the future, they would provide conclusive evidence of a
partial disruption of planets around WDs.

\section{Acknowledgments}

We would like to thank the anonymous referee for helpful
suggestions that led to a significant improvement of our work.
This study was supported by the Natural Science Foundation of
Xinjiang Uygur Autonomous Region (Nos. 2022D01A363, 2023D01E20),
the National Natural Science Foundation of China (Grant Nos. 12288102,
12033001, 12273028, 12041304, 12233002, 12041306), the Major
Science and Technology Program of Xinjiang Uygur Autonomous
Region (Nos. 2022A03013-1, 2022A03013-3), the National
SKA Program of China No. 2020SKA0120300, the National Key
R\&D Program of China (2021YFA0718500), the Youth Innovations and
Talents Project of Shandong Provincial Colleges and Universities
(Grant No. 201909118), the Tianshan Talents Training Program
(2023TSYCTD0013). YFH acknowledges the support from the Xinjiang
Tianchi Program. AK acknowledges the support from the Tianchi
Talents Project of Xinjiang Uygur Autonomous Region and the
special research assistance project of the Chinese Academy of
Sciences (CAS). This work was also supported by the Operation,
Maintenance and Upgrading Fund for Astronomical Telescopes and
Facility Instruments, budgeted from the Ministry of Finance of
China (MOF) and administrated by the CAS, the Urumqi Nanshan
Astronomy and Deep Space Exploration Observation and Research
Station of Xinjiang (XJYWZ2303).

\software{SciPy \citep{Virtanen2020}, Matplotlib \citep{Hunter2007}, NumPy \citep{Walt2011}}.


\nocite{*}
\bibliographystyle{aasjournal}
\bibliography{reference}

\begin{thebibliography}{}
\expandafter\ifx\csname natexlab\endcsname\relax\def\natexlab#1{#1}\fi
\providecommand{\url}[1]{\href{#1}{#1}}
\providecommand{\dodoi}[1]{doi:~\href{http://doi.org/#1}{\nolinkurl{#1}}}
\providecommand{\doeprint}[1]{\href{http://ascl.net/#1}{\nolinkurl{http://ascl.net/#1}}}
\providecommand{\doarXiv}[1]{\href{https://arxiv.org/abs/#1}{\nolinkurl{https://arxiv.org/abs/#1}}}

\bibitem[{{Adams} \& {Laughlin}(2003)}]{Adams2003Icar}
{Adams}, F.~C., \& {Laughlin}, G. 2003, \icarus, 163, 290,
  \dodoi{10.1016/S0019-1035(03)00081-2}

\bibitem[{{Antoniadou} \& {Veras}(2016)}]{Antoniadou2016MNRAS}
{Antoniadou}, K.~I., \& {Veras}, D. 2016, \mnras, 463, 4108,
  \dodoi{10.1093/mnras/stw2264}

\bibitem[{{Antonini} {et~al.}(2016){Antonini}, {Chatterjee}, {Rodriguez},
  {Morscher}, {Pattabiraman}, {Kalogera}, \& {Rasio}}]{Antonini2016ApJ}
{Antonini}, F., {Chatterjee}, S., {Rodriguez}, C.~L., {et~al.} 2016, \apj, 816,
  65, \dodoi{10.3847/0004-637X/816/2/65}

\bibitem[{{Antonini} {et~al.}(2014){Antonini}, {Murray}, \&
  {Mikkola}}]{Antonini2014ApJ}
{Antonini}, F., {Murray}, N., \& {Mikkola}, S. 2014, \apj, 781, 45,
  \dodoi{10.1088/0004-637X/781/1/45}

\bibitem[{{Antonini} \& {Perets}(2012)}]{Antonini2012ApJ}
{Antonini}, F., \& {Perets}, H.~B. 2012, \apj, 757, 27,
  \dodoi{10.1088/0004-637X/757/1/27}

\bibitem[{{Armstrong} {et~al.}(2020){Armstrong}, {Lopez}, {Adibekyan}, {Booth},
  {Bryant}, {Collins}, {Deleuil}, {Emsenhuber}, {Huang}, {King}, {Lillo-Box},
  {Lissauer}, {Matthews}, {Mousis}, {Nielsen}, {Osborn}, {Otegi}, {Santos},
  {Sousa}, {Stassun}, {Veras}, {Ziegler}, {Acton}, {Almenara}, {Anderson},
  {Barrado}, {Barros}, {Bayliss}, {Belardi}, {Bouchy}, {Brice{\~n}o}, {Brogi},
  {Brown}, {Burleigh}, {Casewell}, {Chaushev}, {Ciardi}, {Collins},
  {Col{\'o}n}, {Cooke}, {Crossfield}, {D{\'\i}az}, {Delgado Mena}, {Demangeon},
  {Dorn}, {Dumusque}, {Eigm{\"u}ller}, {Fausnaugh}, {Figueira}, {Gan},
  {Gandhi}, {Gill}, {Gonzales}, {Goad}, {G{\"u}nther}, {Helled}, {Hojjatpanah},
  {Howell}, {Jackman}, {Jenkins}, {Jenkins}, {Jensen}, {Kennedy}, {Latham},
  {Law}, {Lendl}, {Lozovsky}, {Mann}, {Moyano}, {McCormac}, {Meru},
  {Mordasini}, {Osborn}, {Pollacco}, {Queloz}, {Raynard}, {Ricker}, {Rowden},
  {Santerne}, {Schlieder}, {Seager}, {Sha}, {Tan}, {Tilbrook}, {Ting}, {Udry},
  {Vanderspek}, {Watson}, {West}, {Wilson}, {Winn}, {Wheatley}, {Villasenor},
  {Vines}, \& {Zhan}}]{Armstrong2020Natur}
{Armstrong}, D.~J., {Lopez}, T.~A., {Adibekyan}, V., {et~al.} 2020, \nat, 583,
  39, \dodoi{10.1038/s41586-020-2421-7}

\bibitem[{Bareither {et~al.}(2008)Bareither, Edil, Benson, \&
  Mickelson}]{Bareither2008}
Bareither, C.~A., Edil, T.~B., Benson, C.~H., \& Mickelson, D.~M. 2008, Journal
  of Geotechnical and Geoenvironmental Engineering, 134, 1476,
  \dodoi{10.1061/(ASCE)1090-0241(2008)134:10(1476)}

\bibitem[{{Bhaskar} {et~al.}(2021){Bhaskar}, {Li}, {Hadden}, {Payne}, \&
  {Holman}}]{Bhaskar2021AJ}
{Bhaskar}, H., {Li}, G., {Hadden}, S., {Payne}, M.~J., \& {Holman}, M.~J. 2021,
  \aj, 161, 48, \dodoi{10.3847/1538-3881/abcbfc}

\bibitem[{{Blackman} {et~al.}(2021){Blackman}, {Beaulieu}, {Bennett},
  {Danielski}, {Alard}, {Cole}, {Vandorou}, {Ranc}, {Terry}, {Bhattacharya},
  {Bond}, {Bachelet}, {Veras}, {Koshimoto}, {Batista}, \&
  {Marquette}}]{Blackman2021Natur}
{Blackman}, J.~W., {Beaulieu}, J.~P., {Bennett}, D.~P., {et~al.} 2021, \nat,
  598, 272, \dodoi{10.1038/s41586-021-03869-6}

\bibitem[{{Blouin} \& {Xu}(2022)}]{Blouin2022MNRAS}
{Blouin}, S., \& {Xu}, S. 2022, \mnras, 510, 1059,
  \dodoi{10.1093/mnras/stab3446}

\bibitem[{{Bode} \& {Wegg}(2014)}]{Bode2014MNRAS}
{Bode}, J.~N., \& {Wegg}, C. 2014, \mnras, 438, 573,
  \dodoi{10.1093/mnras/stt2227}

\bibitem[{{Bonsor} {et~al.}(2011){Bonsor}, {Mustill}, \&
  {Wyatt}}]{Bonsor2011MNRAS}
{Bonsor}, A., {Mustill}, A.~J., \& {Wyatt}, M.~C. 2011, \mnras, 414, 930,
  \dodoi{10.1111/j.1365-2966.2011.18524.x}

\bibitem[{{Bonsor} \& {Veras}(2015)}]{Bonsor2015MNRAS}
{Bonsor}, A., \& {Veras}, D. 2015, \mnras, 454, 53,
  \dodoi{10.1093/mnras/stv1913}

\bibitem[{{Brouwers} {et~al.}(2022){Brouwers}, {Bonsor}, \&
  {Malamud}}]{Brouwers2022MNRAS}
{Brouwers}, M.~G., {Bonsor}, A., \& {Malamud}, U. 2022, \mnras, 509, 2404,
  \dodoi{10.1093/mnras/stab3009}

\bibitem[{{Budaj} {et~al.}(2022){Budaj}, {Maliuk}, \& {Hubeny}}]{Budaj2022AA}
{Budaj}, J., {Maliuk}, A., \& {Hubeny}, I. 2022, \aap, 660, A72,
  \dodoi{10.1051/0004-6361/202141924}

\bibitem[{{Carrera} {et~al.}(2019){Carrera}, {Raymond}, \&
  {Davies}}]{Carrera2019AA}
{Carrera}, D., {Raymond}, S.~N., \& {Davies}, M.~B. 2019, \aap, 629, L7,
  \dodoi{10.1051/0004-6361/201935744}

\bibitem[{{Cassan} {et~al.}(2012){Cassan}, {Kubas}, {Beaulieu}, {Dominik},
  {Horne}, {Greenhill}, {Wambsganss}, {Menzies}, {Williams}, {J{\o}rgensen},
  {Udalski}, {Bennett}, {Albrow}, {Batista}, {Brillant}, {Caldwell}, {Cole},
  {Coutures}, {Cook}, {Dieters}, {Dominis Prester}, {Donatowicz}, {Fouqu{\'e}},
  {Hill}, {Kains}, {Kane}, {Marquette}, {Martin}, {Pollard}, {Sahu}, {Vinter},
  {Warren}, {Watson}, {Zub}, {Sumi}, {Szyma{\'n}ski}, {Kubiak}, {Poleski},
  {Soszynski}, {Ulaczyk}, {Pietrzy{\'n}ski}, \&
  {Wyrzykowski}}]{Cassan2012Natur}
{Cassan}, A., {Kubas}, D., {Beaulieu}, J.~P., {et~al.} 2012, \nat, 481, 167,
  \dodoi{10.1038/nature10684}

\bibitem[{{Chambers}(1999)}]{Chambers1999MNRAS}
{Chambers}, J.~E. 1999, \mnras, 304, 793,
  \dodoi{10.1046/j.1365-8711.1999.02379.x}

\bibitem[{{Chen} {et~al.}(2019){Chen}, {Zhou}, {Xie}, {Yang}, {Zhang}, {Liu},
  {Liang}, {Yu}, \& {Yang}}]{Chen2019NatAs}
{Chen}, D.-C., {Zhou}, J.-L., {Xie}, J.-W., {et~al.} 2019, Nature Astronomy, 3,
  69, \dodoi{10.1038/s41550-018-0609-7}

\bibitem[{{Chu} {et~al.}(2021){Chu}, {Toal{\'a}}, {Guerrero}, {Bauer},
  {Bilikova}, \& {Gruendl}}]{Chu2021ApJ}
{Chu}, Y.-H., {Toal{\'a}}, J.~A., {Guerrero}, M.~A., {et~al.} 2021, \apj, 910,
  119, \dodoi{10.3847/1538-4357/abe5a5}

\bibitem[{{Coughlin} \& {Nixon}(2022)}]{Coughlin2022MNRAS}
{Coughlin}, E.~R., \& {Nixon}, C.~J. 2022, \mnras, 517, L26,
  \dodoi{10.1093/mnrasl/slac106}

\bibitem[{{Cummings} {et~al.}(2018){Cummings}, {Kalirai}, {Tremblay},
  {Ramirez-Ruiz}, \& {Choi}}]{Cummings2018ApJ}
{Cummings}, J.~D., {Kalirai}, J.~S., {Tremblay}, P.~E., {Ramirez-Ruiz}, E., \&
  {Choi}, J. 2018, \apj, 866, 21, \dodoi{10.3847/1538-4357/aadfd6}

\bibitem[{{Cunningham} {et~al.}(2022){Cunningham}, {Wheatley}, {Tremblay},
  {G{\"a}nsicke}, {King}, {Toloza}, \& {Veras}}]{Cunningham2022Natur}
{Cunningham}, T., {Wheatley}, P.~J., {Tremblay}, P.-E., {et~al.} 2022, \nat,
  602, 219, \dodoi{10.1038/s41586-021-04300-w}

\bibitem[{{Debes} \& {Sigurdsson}(2002)}]{Debes2002ApJ}
{Debes}, J.~H., \& {Sigurdsson}, S. 2002, \apj, 572, 556,
  \dodoi{10.1086/340291}

\bibitem[{{Debes} {et~al.}(2012){Debes}, {Walsh}, \& {Stark}}]{Debes2012ApJ}
{Debes}, J.~H., {Walsh}, K.~J., \& {Stark}, C. 2012, \apj, 747, 148,
  \dodoi{10.1088/0004-637X/747/2/148}

\bibitem[{{Duvvuri} {et~al.}(2020){Duvvuri}, {Redfield}, \&
  {Veras}}]{Duvvuri2020ApJ...893..166D}
{Duvvuri}, G.~M., {Redfield}, S., \& {Veras}, D. 2020, \apj, 893, 166,
  \dodoi{10.3847/1538-4357/ab7fa0}

\bibitem[{{Eggleton} \& {Kiseleva}(1995)}]{Eggleton1995ApJ}
{Eggleton}, P., \& {Kiseleva}, L. 1995, \apj, 455, 640, \dodoi{10.1086/176611}

\bibitem[{{Estrada-Dorado} {et~al.}(2023){Estrada-Dorado}, {Guerrero},
  {Toal{\'a}}, {Chu}, {Lora}, \&
  {Rodr{\'\i}guez-L{\'o}pez}}]{Estrada-Dorado2023ApJ}
{Estrada-Dorado}, S., {Guerrero}, M.~A., {Toal{\'a}}, J.~A., {et~al.} 2023,
  \apjl, 944, L46, \dodoi{10.3847/2041-8213/acba7e}

\bibitem[{{Faber} {et~al.}(2005){Faber}, {Rasio}, \& {Willems}}]{Faber2005Icar}
{Faber}, J.~A., {Rasio}, F.~A., \& {Willems}, B. 2005, \icarus, 175, 248,
  \dodoi{10.1016/j.icarus.2004.10.021}

\bibitem[{{Farihi}(2016)}]{Farihi2016NewAR}
{Farihi}, J. 2016, \nar, 71, 9, \dodoi{10.1016/j.newar.2016.03.001}

\bibitem[{{Farihi} {et~al.}(2012){Farihi}, {G{\"a}nsicke}, {Wyatt}, {Girven},
  {Pringle}, \& {King}}]{Farihi2012MNRAS}
{Farihi}, J., {G{\"a}nsicke}, B.~T., {Wyatt}, M.~C., {et~al.} 2012, \mnras,
  424, 464, \dodoi{10.1111/j.1365-2966.2012.21215.x}

\bibitem[{{Farihi} {et~al.}(2022){Farihi}, {Hermes}, {Marsh}, {Mustill},
  {Wyatt}, {Guidry}, {Wilson}, {Redfield}, {Izquierdo}, {Toloza},
  {G{\"a}nsicke}, {Aungwerojwit}, {Kaewmanee}, {Dhillon}, \&
  {Swan}}]{Farihi2022MNRAS}
{Farihi}, J., {Hermes}, J.~J., {Marsh}, T.~R., {et~al.} 2022, \mnras, 511,
  1647, \dodoi{10.1093/mnras/stab3475}

\bibitem[{{Frewen} \& {Hansen}(2014)}]{Frewen2014MNRAS}
{Frewen}, S.~F.~N., \& {Hansen}, B.~M.~S. 2014, \mnras, 439, 2442,
  \dodoi{10.1093/mnras/stu097}

\bibitem[{{G{\"a}nsicke} {et~al.}(2019){G{\"a}nsicke}, {Schreiber}, {Toloza},
  {Gentile Fusillo}, {Koester}, \& {Manser}}]{Gansicke2019Natur}
{G{\"a}nsicke}, B.~T., {Schreiber}, M.~R., {Toloza}, O., {et~al.} 2019, \nat,
  576, 61, \dodoi{10.1038/s41586-019-1789-8}

\bibitem[{{Geng} {et~al.}(2015){Geng}, {Huang}, \& {Lu}}]{Geng2015ApJ_a}
{Geng}, J.~J., {Huang}, Y.~F., \& {Lu}, T. 2015, \apj, 804, 21,
  \dodoi{10.1088/0004-637X/804/1/21}

\bibitem[{{Girven} {et~al.}(2012){Girven}, {Brinkworth}, {Farihi},
  {G{\"a}nsicke}, {Hoard}, {Marsh}, \& {Koester}}]{Girven2012ApJ}
{Girven}, J., {Brinkworth}, C.~S., {Farihi}, J., {et~al.} 2012, \apj, 749, 154,
  \dodoi{10.1088/0004-637X/749/2/154}

\bibitem[{{Goffo} {et~al.}(2023){Goffo}, {Gandolfi}, {Egger}, {Mustill},
  {Albrecht}, {Hirano}, {Kochukhov}, {Astudillo-Defru}, {Barragan}, {Serrano},
  {Hatzes}, {Alibert}, {Guenther}, {Dai}, {Lam}, {Csizmadia}, {Smith},
  {Fossati}, {Luque}, {Rodler}, {Winther}, {R{\o}rsted}, {Alarcon}, {Bonfils},
  {Cochran}, {Deeg}, {Jenkins}, {Korth}, {Livingston}, {Meech}, {Murgas},
  {Orell-Miquel}, {Osborne}, {Palle}, {Persson}, {Redfield}, {Ricker},
  {Seager}, {Vanderspek}, {Van Eylen}, \& {Winn}}]{Goffo2023ApJ}
{Goffo}, E., {Gandolfi}, D., {Egger}, J.~A., {et~al.} 2023, \apjl, 955, L3,
  \dodoi{10.3847/2041-8213/ace0c7}

\bibitem[{{Goulinski} \& {Ribak}(2018)}]{Goulinski2018MNRAS}
{Goulinski}, N., \& {Ribak}, E.~N. 2018, \mnras, 473, 1589,
  \dodoi{10.1093/mnras/stx2506}

\bibitem[{{Granvik} {et~al.}(2016){Granvik}, {Morbidelli}, {Jedicke}, {Bolin},
  {Bottke}, {Beshore}, {Vokrouhlick{\'y}}, {Delb{\`o}}, \&
  {Michel}}]{Granvik2016Natur}
{Granvik}, M., {Morbidelli}, A., {Jedicke}, R., {et~al.} 2016, \nat, 530, 303,
  \dodoi{10.1038/nature16934}

\bibitem[{{Grishin} \& {Veras}(2019)}]{Grishin2019MNRAS}
{Grishin}, E., \& {Veras}, D. 2019, \mnras, 489, 168,
  \dodoi{10.1093/mnras/stz2148}

\bibitem[{{Guidry} {et~al.}(2021){Guidry}, {Vanderbosch}, {Hermes}, {Barlow},
  {Lopez}, {Boudreaux}, {Corcoran}, {Bell}, {Montgomery}, {Heintz},
  {Castanheira}, {Reding}, {Dunlap}, {Winget}, {Winget}, \&
  {Kuehne}}]{Guidry2021ApJ}
{Guidry}, J.~A., {Vanderbosch}, Z.~P., {Hermes}, J.~J., {et~al.} 2021, \apj,
  912, 125, \dodoi{10.3847/1538-4357/abee68}

\bibitem[{{Guillochon} \& {Ramirez-Ruiz}(2013)}]{Guillochon2013ApJ}
{Guillochon}, J., \& {Ramirez-Ruiz}, E. 2013, \apj, 767, 25,
  \dodoi{10.1088/0004-637X/767/1/25}

\bibitem[{{Guillochon} {et~al.}(2011){Guillochon}, {Ramirez-Ruiz}, \&
  {Lin}}]{Guillochon2011ApJ}
{Guillochon}, J., {Ramirez-Ruiz}, E., \& {Lin}, D. 2011, \apj, 732, 74,
  \dodoi{10.1088/0004-637X/732/2/74}

\bibitem[{{Hamers} {et~al.}(2022){Hamers}, {Perets}, {Thompson}, \&
  {Neunteufel}}]{Hamers2022ApJ}
{Hamers}, A.~S., {Perets}, H.~B., {Thompson}, T.~A., \& {Neunteufel}, P. 2022,
  \apj, 925, 178, \dodoi{10.3847/1538-4357/ac400b}

\bibitem[{{Hamers} \& {Portegies Zwart}(2016)}]{Hamers2016MNRAS}
{Hamers}, A.~S., \& {Portegies Zwart}, S.~F. 2016, \mnras, 462, L84,
  \dodoi{10.1093/mnrasl/slw134}

\bibitem[{{Harrison} {et~al.}(2021){Harrison}, {Bonsor}, {Kama}, {Buchan},
  {Blouin}, \& {Koester}}]{Harrison2021MNRAS}
{Harrison}, J. H.~D., {Bonsor}, A., {Kama}, M., {et~al.} 2021, \mnras, 504,
  2853, \dodoi{10.1093/mnras/stab736}

\bibitem[{{Hartmann} {et~al.}(2011){Hartmann}, {Nagel}, {Rauch}, \&
  {Werner}}]{Hartmann2011AA}
{Hartmann}, S., {Nagel}, T., {Rauch}, T., \& {Werner}, K. 2011, \aap, 530, A7,
  \dodoi{10.1051/0004-6361/201116625}

\bibitem[{{He} \& {Petrovich}(2018)}]{He2018MNRAS}
{He}, M.~Y., \& {Petrovich}, C. 2018, \mnras, 474, 20,
  \dodoi{10.1093/mnras/stx2718}

\bibitem[{{Hills}(1975)}]{Hills1975Natur}
{Hills}, J.~G. 1975, \nat, 254, 295, \dodoi{10.1038/254295a0}

\bibitem[{{Hogg} {et~al.}(2021){Hogg}, {Cutter}, \& {Wynn}}]{Hogg2021MNRAS}
{Hogg}, M.~A., {Cutter}, R., \& {Wynn}, G.~A. 2021, \mnras, 500, 2986,
  \dodoi{10.1093/mnras/staa3316}

\bibitem[{{Hollands} {et~al.}(2018){Hollands}, {G{\"a}nsicke}, \&
  {Koester}}]{Hollands2018MNRAS}
{Hollands}, M.~A., {G{\"a}nsicke}, B.~T., \& {Koester}, D. 2018, \mnras, 477,
  93, \dodoi{10.1093/mnras/sty592}

\bibitem[{{Huang} \& {Yu}(2017)}]{Huang2017ApJ}
{Huang}, Y.~F., \& {Yu}, Y.~B. 2017, \apj, 848, 115,
  \dodoi{10.3847/1538-4357/aa8b63}

\bibitem[{{Hunter}(2007)}]{Hunter2007}
{Hunter}, J.~D. 2007, Computing in Science and Engineering, 9, 90,
  \dodoi{10.1109/MCSE.2007.55}

\bibitem[{{Jura}(2003)}]{Jura2003ApJ}
{Jura}, M. 2003, \apjl, 584, L91, \dodoi{10.1086/374036}

\bibitem[{{Katz} \& {Dong}(2012)}]{Katz2012}
{Katz}, B., \& {Dong}, S. 2012, arXiv e-prints, arXiv:1211.4584.
\newblock \doarXiv{1211.4584}

\bibitem[{{Klein} {et~al.}(2021){Klein}, {Doyle}, {Zuckerman}, {Dufour},
  {Blouin}, {Melis}, {Weinberger}, \& {Young}}]{Klein2021ApJ}
{Klein}, B.~L., {Doyle}, A.~E., {Zuckerman}, B., {et~al.} 2021, \apj, 914, 61,
  \dodoi{10.3847/1538-4357/abe40b}

\bibitem[{{Kleinman} {et~al.}(2013){Kleinman}, {Kepler}, {Koester}, {Pelisoli},
  {Pe{\c{c}}anha}, {Nitta}, {Costa}, {Krzesinski}, {Dufour}, {Lachapelle},
  {Bergeron}, {Yip}, {Harris}, {Eisenstein}, {Althaus}, \&
  {C{\'o}rsico}}]{Kleinman2013ApJS}
{Kleinman}, S.~J., {Kepler}, S.~O., {Koester}, D., {et~al.} 2013, \apjs, 204,
  5, \dodoi{10.1088/0067-0049/204/1/5}

\bibitem[{{Koester}(2009)}]{Koester2009AA}
{Koester}, D. 2009, \aap, 498, 517, \dodoi{10.1051/0004-6361/200811468}

\bibitem[{{Koester} {et~al.}(2014){Koester}, {G{\"a}nsicke}, \&
  {Farihi}}]{Koester2014AA}
{Koester}, D., {G{\"a}nsicke}, B.~T., \& {Farihi}, J. 2014, \aap, 566, A34,
  \dodoi{10.1051/0004-6361/201423691}

\bibitem[{{Kozai}(1962)}]{Kozai1962}
{Kozai}, Y. 1962, \aj, 67, 591, \dodoi{10.1086/108790}

\bibitem[{{Kratter} \& {Perets}(2012)}]{Kratter2012ApJ...753...91K}
{Kratter}, K.~M., \& {Perets}, H.~B. 2012, \apj, 753, 91,
  \dodoi{10.1088/0004-637X/753/1/91}

\bibitem[{{Kremer} {et~al.}(2019){Kremer}, {D'Orazio}, {Samsing}, {Chatterjee},
  \& {Rasio}}]{Kremer2019ApJ}
{Kremer}, K., {D'Orazio}, D.~J., {Samsing}, J., {Chatterjee}, S., \& {Rasio},
  F.~A. 2019, \apj, 885, 2, \dodoi{10.3847/1538-4357/ab44d1}

\bibitem[{{Kuerban} {et~al.}(2019){Kuerban}, {Geng}, \&
  {Huang}}]{Kuerban2019AIPC}
{Kuerban}, A., {Geng}, J.-J., \& {Huang}, Y.-F. 2019, in American Institute of
  Physics Conference Series, Vol. 2127, Xiamen-CUSTIPEN Workshop on the
  Equation of State of Dense Neutron-Rich Matter in the Era of Gravitational
  Wave Astronomy, 020027, \dodoi{10.1063/1.5117817}

\bibitem[{{Kuerban} {et~al.}(2020){Kuerban}, {Geng}, {Huang}, {Zong}, \&
  {Gong}}]{Kuerban2020ApJ}
{Kuerban}, A., {Geng}, J.-J., {Huang}, Y.-F., {Zong}, H.-S., \& {Gong}, H.
  2020, \apj, 890, 41, \dodoi{10.3847/1538-4357/ab698b}

\bibitem[{{Kunitomo} {et~al.}(2011){Kunitomo}, {Ikoma}, {Sato}, {Katsuta}, \&
  {Ida}}]{Kunitomo2011ApJ}
{Kunitomo}, M., {Ikoma}, M., {Sato}, B., {Katsuta}, Y., \& {Ida}, S. 2011,
  \apj, 737, 66, \dodoi{10.1088/0004-637X/737/2/66}

\bibitem[{{Kurban} {et~al.}(2023){Kurban}, {Zhou}, {Wang}, {Huang}, {Wang}, \&
  {Nurmamat}}]{Kurban2023MNRAS}
{Kurban}, A., {Zhou}, X., {Wang}, N., {et~al.} 2023, \mnras, 522, 4265,
  \dodoi{10.1093/mnras/stad1260}

\bibitem[{{Kurban} {et~al.}(2024){Kurban}, {Zhou}, {Wang}, {Huang}, {Wang}, \&
  {Nurmamat}}]{Kurban2024AA}
---. 2024, \aap, 686, A87, \dodoi{10.1051/0004-6361/202347828}

\bibitem[{{Kurban} {et~al.}(2022){Kurban}, {Huang}, {Geng}, {Li}, {Xu}, {Wang},
  {Zhou}, {Esamdin}, \& {Wang}}]{Kurban2022ApJ}
{Kurban}, A., {Huang}, Y.-F., {Geng}, J.-J., {et~al.} 2022, \apj, 928, 94,
  \dodoi{10.3847/1538-4357/ac558f}

\bibitem[{{Lam} {et~al.}(2021){Lam}, {Csizmadia}, {Astudillo-Defru}, {Bonfils},
  {Gandolfi}, {Padovan}, {Esposito}, {Hellier}, {Hirano}, {Livingston},
  {Murgas}, {Smith}, {Collins}, {Mathur}, {Garcia}, {Howell}, {Santos}, {Dai},
  {Ricker}, {Vanderspek}, {Latham}, {Seager}, {Winn}, {Jenkins}, {Albrecht},
  {Almenara}, {Artigau}, {Barrag{\'a}n}, {Bouchy}, {Cabrera}, {Charbonneau},
  {Chaturvedi}, {Chaushev}, {Christiansen}, {Cochran}, {De Meideiros},
  {Delfosse}, {D{\'\i}az}, {Doyon}, {Eigm{\"u}ller}, {Figueira}, {Forveille},
  {Fridlund}, {Gaisn{\'e}}, {Goffo}, {Georgieva}, {Grziwa}, {Guenther},
  {Hatzes}, {Johnson}, {Kab{\'a}th}, {Knudstrup}, {Korth}, {Lewin}, {Lissauer},
  {Lovis}, {Luque}, {Melo}, {Morgan}, {Morris}, {Mayor}, {Narita}, {Osborne},
  {Palle}, {Pepe}, {Persson}, {Quinn}, {Rauer}, {Redfield}, {Schlieder},
  {S{\'e}gransan}, {Serrano}, {Smith}, {{\v{S}}ubjak}, {Twicken}, {Udry}, {Van
  Eylen}, \& {Vezie}}]{Lam2021Sci}
{Lam}, K. W.~F., {Csizmadia}, S., {Astudillo-Defru}, N., {et~al.} 2021,
  Science, 374, 1271, \dodoi{10.1126/science.aay3253}

\bibitem[{{Law-Smith} {et~al.}(2020){Law-Smith}, {Coulter}, {Guillochon},
  {Mockler}, \& {Ramirez-Ruiz}}]{Law-Smith2020ApJ}
{Law-Smith}, J. A.~P., {Coulter}, D.~A., {Guillochon}, J., {Mockler}, B., \&
  {Ramirez-Ruiz}, E. 2020, \apj, 905, 141, \dodoi{10.3847/1538-4357/abc489}

\bibitem[{{Li} {et~al.}(2021){Li}, {Mustill}, \& {Davies}}]{Li2021MNRAS}
{Li}, D., {Mustill}, A.~J., \& {Davies}, M.~B. 2021, \mnras, 508, 5671,
  \dodoi{10.1093/mnras/stab2949}

\bibitem[{{Lidov}(1962)}]{Lidov1962}
{Lidov}, M.~L. 1962, Planetary and Space Science, 9, 719,
  \dodoi{10.1016/0032-0633(62)90129-0}

\bibitem[{{Liu} {et~al.}(2013){Liu}, {Guillochon}, {Lin}, \&
  {Ramirez-Ruiz}}]{Liu2013ApJ}
{Liu}, S.-F., {Guillochon}, J., {Lin}, D. N.~C., \& {Ramirez-Ruiz}, E. 2013,
  \apj, 762, 37, \dodoi{10.1088/0004-637X/762/1/37}

\bibitem[{{Luhman} {et~al.}(2011){Luhman}, {Burgasser}, \&
  {Bochanski}}]{Luhman2011ApJ}
{Luhman}, K.~L., {Burgasser}, A.~J., \& {Bochanski}, J.~J. 2011, \apjl, 730,
  L9, \dodoi{10.1088/2041-8205/730/1/L9}

\bibitem[{{Malamud} {et~al.}(2021){Malamud}, {Grishin}, \&
  {Brouwers}}]{Malamud2021MNRAS}
{Malamud}, U., {Grishin}, E., \& {Brouwers}, M. 2021, \mnras, 501, 3806,
  \dodoi{10.1093/mnras/staa3940}

\bibitem[{{Malamud} \& {Perets}(2020{\natexlab{a}})}]{Malamud2020a}
{Malamud}, U., \& {Perets}, H.~B. 2020{\natexlab{a}}, \mnras, 492, 5561,
  \dodoi{10.1093/mnras/staa142}

\bibitem[{{Malamud} \& {Perets}(2020{\natexlab{b}})}]{Malamud2020b}
---. 2020{\natexlab{b}}, \mnras, 493, 698, \dodoi{10.1093/mnras/staa143}

\bibitem[{{Malavolta} {et~al.}(2018){Malavolta}, {Mayo}, {Louden}, {Rajpaul},
  {Bonomo}, {Buchhave}, {Kreidberg}, {Kristiansen}, {Lopez-Morales}, {Mortier},
  {Vanderburg}, {Coffinet}, {Ehrenreich}, {Lovis}, {Bouchy}, {Charbonneau},
  {Ciardi}, {Collier Cameron}, {Cosentino}, {Crossfield}, {Damasso},
  {Dressing}, {Dumusque}, {Everett}, {Figueira}, {Fiorenzano}, {Gonzales},
  {Haywood}, {Harutyunyan}, {Hirsch}, {Howell}, {Johnson}, {Latham}, {Lopez},
  {Mayor}, {Micela}, {Molinari}, {Nascimbeni}, {Pepe}, {Phillips}, {Piotto},
  {Rice}, {Sasselov}, {S{\'e}gransan}, {Sozzetti}, {Udry}, \&
  {Watson}}]{Malavolta2018AJ}
{Malavolta}, L., {Mayo}, A.~W., {Louden}, T., {et~al.} 2018, \aj, 155, 107,
  \dodoi{10.3847/1538-3881/aaa5b5}

\bibitem[{{Maldonado} {et~al.}(2022){Maldonado}, {Villaver}, {Mustill}, \&
  {Ch{\'a}vez}}]{Maldonado2022MNRAS}
{Maldonado}, R.~F., {Villaver}, E., {Mustill}, A.~J., \& {Ch{\'a}vez}, M. 2022,
  \mnras, 512, 104, \dodoi{10.1093/mnras/stac481}

\bibitem[{{Maldonado} {et~al.}(2020{\natexlab{a}}){Maldonado}, {Villaver},
  {Mustill}, {Chavez}, \& {Bertone}}]{Maldonado2020aMNRAS}
{Maldonado}, R.~F., {Villaver}, E., {Mustill}, A.~J., {Chavez}, M., \&
  {Bertone}, E. 2020{\natexlab{a}}, \mnras, 497, 4091,
  \dodoi{10.1093/mnras/staa2237}

\bibitem[{{Maldonado} {et~al.}(2020{\natexlab{b}}){Maldonado}, {Villaver},
  {Mustill}, {Chavez}, \& {Bertone}}]{Maldonado2020bMNRAS}
---. 2020{\natexlab{b}}, \mnras, 499, 1854, \dodoi{10.1093/mnras/staa2946}

\bibitem[{{Maldonado} {et~al.}(2021){Maldonado}, {Villaver}, {Mustill},
  {Ch{\'a}vez}, \& {Bertone}}]{Maldonado2021MNRAS}
{Maldonado}, R.~F., {Villaver}, E., {Mustill}, A.~J., {Ch{\'a}vez}, M., \&
  {Bertone}, E. 2021, \mnras, 501, L43, \dodoi{10.1093/mnrasl/slaa193}

\bibitem[{{Manser} {et~al.}(2019){Manser}, {G{\"a}nsicke}, {Eggl}, {Hollands},
  {Izquierdo}, {Koester}, {Landstreet}, {Lyra}, {Marsh}, {Meru}, {Mustill},
  {Rodr{\'\i}guez-Gil}, {Toloza}, {Veras}, {Wilson}, {Burleigh}, {Davies},
  {Farihi}, {Gentile Fusillo}, {de Martino}, {Parsons}, {Quirrenbach}, {Raddi},
  {Reffert}, {Del Santo}, {Schreiber}, {Silvotti}, {Toonen}, {Villaver},
  {Wyatt}, {Xu}, \& {Portegies Zwart}}]{Manser2019Sci}
{Manser}, C.~J., {G{\"a}nsicke}, B.~T., {Eggl}, S., {et~al.} 2019, Science,
  364, 66, \dodoi{10.1126/science.aat5330}

\bibitem[{{Mardling} \& {Aarseth}(2001)}]{Mardling2001MNRAS}
{Mardling}, R.~A., \& {Aarseth}, S.~J. 2001, \mnras, 321, 398,
  \dodoi{10.1046/j.1365-8711.2001.03974.x}

\bibitem[{{Metzger} {et~al.}(2012){Metzger}, {Rafikov}, \&
  {Bochkarev}}]{Metzger2012MNRAS}
{Metzger}, B.~D., {Rafikov}, R.~R., \& {Bochkarev}, K.~V. 2012, \mnras, 423,
  505, \dodoi{10.1111/j.1365-2966.2012.20895.x}

\bibitem[{{Mu{\~n}oz} \& {Petrovich}(2020)}]{Diego2020ApJ}
{Mu{\~n}oz}, D.~J., \& {Petrovich}, C. 2020, \apjl, 904, L3,
  \dodoi{10.3847/2041-8213/abc564}

\bibitem[{{Mukai}(2017)}]{Mukai2017PASP}
{Mukai}, K. 2017, \pasp, 129, 062001, \dodoi{10.1088/1538-3873/aa6736}

\bibitem[{{Mustill} {et~al.}(2022){Mustill}, {Davies}, {Blunt}, \&
  {Howard}}]{Mustill2022MNRAS}
{Mustill}, A.~J., {Davies}, M.~B., {Blunt}, S., \& {Howard}, A. 2022, \mnras,
  509, 3616, \dodoi{10.1093/mnras/stab3174}

\bibitem[{{Mustill} {et~al.}(2014){Mustill}, {Veras}, \&
  {Villaver}}]{Mustill2014MNRAS}
{Mustill}, A.~J., {Veras}, D., \& {Villaver}, E. 2014, \mnras, 437, 1404,
  \dodoi{10.1093/mnras/stt1973}

\bibitem[{{Mustill} \& {Villaver}(2012)}]{Mustill2012ApJ}
{Mustill}, A.~J., \& {Villaver}, E. 2012, \apj, 761, 121,
  \dodoi{10.1088/0004-637X/761/2/121}

\bibitem[{{Mustill} {et~al.}(2018){Mustill}, {Villaver}, {Veras},
  {G{\"a}nsicke}, \& {Bonsor}}]{Mustill2018MNRAS}
{Mustill}, A.~J., {Villaver}, E., {Veras}, D., {G{\"a}nsicke}, B.~T., \&
  {Bonsor}, A. 2018, \mnras, 476, 3939, \dodoi{10.1093/mnras/sty446}

\bibitem[{{Nagasawa} \& {Ida}(2011)}]{Nagasawa2011ApJ}
{Nagasawa}, M., \& {Ida}, S. 2011, \apj, 742, 72,
  \dodoi{10.1088/0004-637X/742/2/72}

\bibitem[{{Naoz}(2016)}]{Naoz2016ARAA}
{Naoz}, S. 2016, \araa, 54, 441, \dodoi{10.1146/annurev-astro-081915-023315}

\bibitem[{{Naoz} {et~al.}(2011){Naoz}, {Farr}, {Lithwick}, {Rasio}, \&
  {Teyssandier}}]{Naoz2011Natur}
{Naoz}, S., {Farr}, W.~M., {Lithwick}, Y., {Rasio}, F.~A., \& {Teyssandier}, J.
  2011, \nat, 473, 187, \dodoi{10.1038/nature10076}

\bibitem[{{Naponiello} {et~al.}(2023){Naponiello}, {Mancini}, {Sozzetti},
  {Bonomo}, {Morbidelli}, {Dou}, {Zeng}, {Leinhardt}, {Biazzo}, {Cubillos},
  {Pinamonti}, {Locci}, {Maggio}, {Damasso}, {Lanza}, {Lissauer}, {Collins},
  {Carter}, {Jensen}, {Bignamini}, {Boschin}, {Bouma}, {Ciardi}, {Cosentino},
  {Crossfield}, {Desidera}, {Dumusque}, {Fiorenzano}, {Fukui}, {Giacobbe},
  {Gnilka}, {Ghedina}, {Guilluy}, {Harutyunyan}, {Howell}, {Jenkins}, {Lund},
  {Kielkopf}, {Lester}, {Malavolta}, {Mann}, {Matson}, {Matthews}, {Nardiello},
  {Narita}, {Pace}, {Pagano}, {Palle}, {Pedani}, {Seager}, {Schlieder},
  {Schwarz}, {Shporer}, {Twicken}, {Winn}, {Ziegler}, \&
  {Zingales}}]{Naponiello2023Natur}
{Naponiello}, L., {Mancini}, L., {Sozzetti}, A., {et~al.} 2023, \nat, 622, 255,
  \dodoi{10.1038/s41586-023-06499-2}

\bibitem[{{O'Connor} \& {Lai}(2020)}]{Connor2020MNRAS}
{O'Connor}, C.~E., \& {Lai}, D. 2020, \mnras, 498, 4005,
  \dodoi{10.1093/mnras/staa2645}

\bibitem[{{O'Connor} {et~al.}(2021){O'Connor}, {Liu}, \&
  {Lai}}]{Connor2021MNRAS}
{O'Connor}, C.~E., {Liu}, B., \& {Lai}, D. 2021, \mnras, 501, 507,
  \dodoi{10.1093/mnras/staa3723}

\bibitem[{{O'Connor} {et~al.}(2022){O'Connor}, {Teyssandier}, \&
  {Lai}}]{Connor2022MNRAS}
{O'Connor}, C.~E., {Teyssandier}, J., \& {Lai}, D. 2022, \mnras, 513, 4178,
  \dodoi{10.1093/mnras/stac1189}

\bibitem[{{Paquette} {et~al.}(1986){Paquette}, {Pelletier}, {Fontaine}, \&
  {Michaud}}]{Paquette1986ApJS}
{Paquette}, C., {Pelletier}, C., {Fontaine}, G., \& {Michaud}, G. 1986, \apjs,
  61, 197, \dodoi{10.1086/191112}

\bibitem[{{Perets} \& {Kratter}(2012)}]{Perets2012ApJ}
{Perets}, H.~B., \& {Kratter}, K.~M. 2012, \apj, 760, 99,
  \dodoi{10.1088/0004-637X/760/2/99}

\bibitem[{{Petrovich} \& {Mu{\~n}oz}(2017)}]{Petrovich2017ApJ}
{Petrovich}, C., \& {Mu{\~n}oz}, D.~J. 2017, \apj, 834, 116,
  \dodoi{10.3847/1538-4357/834/2/116}

\bibitem[{{Pohl} \& {Britt}(2020)}]{Pohl2020MPS}
{Pohl}, L., \& {Britt}, D.~T. 2020, \maps, 55, 962, \dodoi{10.1111/maps.13449}

\bibitem[{{Rafikov}(2011{\natexlab{a}})}]{Rafikov2011ApJ}
{Rafikov}, R.~R. 2011{\natexlab{a}}, \apjl, 732, L3,
  \dodoi{10.1088/2041-8205/732/1/L3}

\bibitem[{{Rafikov}(2011{\natexlab{b}})}]{Rafikov2011MNRAS}
---. 2011{\natexlab{b}}, \mnras, 416, L55,
  \dodoi{10.1111/j.1745-3933.2011.01096.x}

\bibitem[{{Rees}(1988)}]{Rees1988Natur}
{Rees}, M.~J. 1988, \nat, 333, 523, \dodoi{10.1038/333523a0}

\bibitem[{{Rossi} {et~al.}(2021){Rossi}, {Stone}, {Law-Smith}, {Macleod},
  {Lodato}, {Dai}, \& {Mandel}}]{Rossi2021SSRv}
{Rossi}, E.~M., {Stone}, N.~C., {Law-Smith}, J.~A.~P., {et~al.} 2021, \ssr,
  217, 40, \dodoi{10.1007/s11214-021-00818-7}

\bibitem[{{Ryu} {et~al.}(2020){Ryu}, {Krolik}, {Piran}, \&
  {Noble}}]{Ryu2020ApJ}
{Ryu}, T., {Krolik}, J., {Piran}, T., \& {Noble}, S.~C. 2020, \apj, 904, 100,
  \dodoi{10.3847/1538-4357/abb3ce}

\bibitem[{{Sato} {et~al.}(2008){Sato}, {Izumiura}, {Toyota}, {Kambe}, {Ikoma},
  {Omiya}, {Masuda}, {Takeda}, {Murata}, {Itoh}, {Ando}, {Yoshida}, {Kokubo},
  \& {Ida}}]{Sato2008PASJ}
{Sato}, B., {Izumiura}, H., {Toyota}, E., {et~al.} 2008, \pasj, 60, 539,
  \dodoi{10.1093/pasj/60.3.539}

\bibitem[{{Seager} {et~al.}(2007){Seager}, {Kuchner}, {Hier-Majumder}, \&
  {Militzer}}]{Seager2007ApJ}
{Seager}, S., {Kuchner}, M., {Hier-Majumder}, C.~A., \& {Militzer}, B. 2007,
  \apj, 669, 1279, \dodoi{10.1086/521346}

\bibitem[{{Smith} {et~al.}(2018){Smith}, {Fratanduono}, {Braun}, {Duffy},
  {Wicks}, {Celliers}, {Ali}, {Fernandez-Pa{\~n}ella}, {Kraus}, {Swift},
  {Collins}, \& {Eggert}}]{Smith2018NatAs}
{Smith}, R.~F., {Fratanduono}, D.~E., {Braun}, D.~G., {et~al.} 2018, Nature
  Astronomy, 2, 452, \dodoi{10.1038/s41550-018-0437-9}

\bibitem[{{Stephan} {et~al.}(2021){Stephan}, {Naoz}, \&
  {Gaudi}}]{Stephan2021ApJ}
{Stephan}, A.~P., {Naoz}, S., \& {Gaudi}, B.~S. 2021, \apj, 922, 4,
  \dodoi{10.3847/1538-4357/ac22a9}

\bibitem[{{Stephan} {et~al.}(2017){Stephan}, {Naoz}, \&
  {Zuckerman}}]{Stephan2017ApJ}
{Stephan}, A.~P., {Naoz}, S., \& {Zuckerman}, B. 2017, \apjl, 844, L16,
  \dodoi{10.3847/2041-8213/aa7cf3}

\bibitem[{{Stock} {et~al.}(2022){Stock}, {Veras}, {Cai}, {Spurzem}, \&
  {Portegies Zwart}}]{Stock2022MNRAS}
{Stock}, K., {Veras}, D., {Cai}, M.~X., {Spurzem}, R., \& {Portegies Zwart}, S.
  2022, \mnras, 512, 2460, \dodoi{10.1093/mnras/stac602}

\bibitem[{{Swan} {et~al.}(2021){Swan}, {Kenyon}, {Farihi}, {Dennihy},
  {G{\"a}nsicke}, {Hermes}, {Melis}, \& {von Hippel}}]{Swan2021MNRAS}
{Swan}, A., {Kenyon}, S.~J., {Farihi}, J., {et~al.} 2021, \mnras, 506, 432,
  \dodoi{10.1093/mnras/stab1738}

\bibitem[{{Thorsett} {et~al.}(1993){Thorsett}, {Arzoumanian}, \&
  {Taylor}}]{Thorsett1993ApJ}
{Thorsett}, S.~E., {Arzoumanian}, Z., \& {Taylor}, J.~H. 1993, \apjl, 412, L33,
  \dodoi{10.1086/186933}

\bibitem[{{Toonen} {et~al.}(2022){Toonen}, {Boekholt}, \& {Portegies
  Zwart}}]{Toonen2022AA}
{Toonen}, S., {Boekholt}, T.~C.~N., \& {Portegies Zwart}, S. 2022, \aap, 661,
  A61, \dodoi{10.1051/0004-6361/202141991}

\bibitem[{{van der Walt} {et~al.}(2011){van der Walt}, {Colbert}, \&
  {Varoquaux}}]{Walt2011}
{van der Walt}, S., {Colbert}, S.~C., \& {Varoquaux}, G. 2011, Computing in
  Science and Engineering, 13, 22, \dodoi{10.1109/MCSE.2011.37}

\bibitem[{{Vanderbosch} {et~al.}(2020){Vanderbosch}, {Hermes}, {Dennihy},
  {Dunlap}, {Izquierdo}, {Tremblay}, {Cho}, {G{\"a}nsicke}, {Toloza}, {Bell},
  {Montgomery}, \& {Winget}}]{Vanderbosch2020ApJ}
{Vanderbosch}, Z., {Hermes}, J.~J., {Dennihy}, E., {et~al.} 2020, \apj, 897,
  171, \dodoi{10.3847/1538-4357/ab9649}

\bibitem[{{Vanderbosch} {et~al.}(2021){Vanderbosch}, {Rappaport}, {Guidry},
  {Gary}, {Blouin}, {Kaye}, {Weinberger}, {Melis}, {Klein}, {Zuckerman},
  {Vanderburg}, {Hermes}, {Hegedus}, {Burleigh}, {Sefako}, {Worters}, \&
  {Heintz}}]{Vanderbosch2021ApJ}
{Vanderbosch}, Z.~P., {Rappaport}, S., {Guidry}, J.~A., {et~al.} 2021, \apj,
  917, 41, \dodoi{10.3847/1538-4357/ac0822}

\bibitem[{{Vanderburg} {et~al.}(2015){Vanderburg}, {Johnson}, {Rappaport},
  {Bieryla}, {Irwin}, {Lewis}, {Kipping}, {Brown}, {Dufour}, {Ciardi}, {Angus},
  {Schaefer}, {Latham}, {Charbonneau}, {Beichman}, {Eastman}, {McCrady},
  {Wittenmyer}, \& {Wright}}]{Vanderburg2015Natur}
{Vanderburg}, A., {Johnson}, J.~A., {Rappaport}, S., {et~al.} 2015, \nat, 526,
  546, \dodoi{10.1038/nature15527}

\bibitem[{{Vanderburg} {et~al.}(2020){Vanderburg}, {Rappaport}, {Xu},
  {Crossfield}, {Becker}, {Gary}, {Murgas}, {Blouin}, {Kaye}, {Palle}, {Melis},
  {Morris}, {Kreidberg}, {Gorjian}, {Morley}, {Mann}, {Parviainen}, {Pearce},
  {Newton}, {Carrillo}, {Zuckerman}, {Nelson}, {Zeimann}, {Brown},
  {Tronsgaard}, {Klein}, {Ricker}, {Vanderspek}, {Latham}, {Seager}, {Winn},
  {Jenkins}, {Adams}, {Benneke}, {Berardo}, {Buchhave}, {Caldwell},
  {Christiansen}, {Collins}, {Col{\'o}n}, {Daylan}, {Doty}, {Doyle},
  {Dragomir}, {Dressing}, {Dufour}, {Fukui}, {Glidden}, {Guerrero}, {Guo},
  {Heng}, {Henriksen}, {Huang}, {Kaltenegger}, {Kane}, {Lewis}, {Lissauer},
  {Morales}, {Narita}, {Pepper}, {Rose}, {Smith}, {Stassun}, \&
  {Yu}}]{Vanderburg2020Natur}
{Vanderburg}, A., {Rappaport}, S.~A., {Xu}, S., {et~al.} 2020, \nat, 585, 363,
  \dodoi{10.1038/s41586-020-2713-y}

\bibitem[{{Veras}(2016{\natexlab{a}})}]{Veras2016MNRAS.463.2958V}
{Veras}, D. 2016{\natexlab{a}}, \mnras, 463, 2958,
  \dodoi{10.1093/mnras/stw2170}

\bibitem[{{Veras}(2016{\natexlab{b}})}]{Veras2016RSOS}
---. 2016{\natexlab{b}}, Royal Society Open Science, 3, 150571,
  \dodoi{10.1098/rsos.150571}

\bibitem[{{Veras}(2020)}]{Veras2020MNRAS.493.4692V}
---. 2020, \mnras, 493, 4692, \dodoi{10.1093/mnras/staa625}

\bibitem[{{Veras}(2021)}]{Veras2021orel.bookE}
---. 2021, in Oxford Research Encyclopedia of Planetary Science, 1,
  \dodoi{10.1093/acrefore/9780190647926.013.238}

\bibitem[{{Veras} {et~al.}(2022){Veras}, {Birader}, \&
  {Zaman}}]{Veras2022MNRAS}
{Veras}, D., {Birader}, Y., \& {Zaman}, U. 2022, \mnras, 510, 3379,
  \dodoi{10.1093/mnras/stab3490}

\bibitem[{{Veras} {et~al.}(2015{\natexlab{a}}){Veras}, {Eggl}, \&
  {G{\"a}nsicke}}]{Veras2015MNRAS.452.1945V}
{Veras}, D., {Eggl}, S., \& {G{\"a}nsicke}, B.~T. 2015{\natexlab{a}}, \mnras,
  452, 1945, \dodoi{10.1093/mnras/stv1417}

\bibitem[{{Veras} \& {Fuller}(2020)}]{Veras2020MNRAS.492.6059V}
{Veras}, D., \& {Fuller}, J. 2020, \mnras, 492, 6059,
  \dodoi{10.1093/mnras/staa309}

\bibitem[{{Veras} {et~al.}(2017){Veras}, {Georgakarakos}, {Dobbs-Dixon}, \&
  {G{\"a}nsicke}}]{Veras2017MNRAS.465.2053V}
{Veras}, D., {Georgakarakos}, N., {Dobbs-Dixon}, I., \& {G{\"a}nsicke}, B.~T.
  2017, \mnras, 465, 2053, \dodoi{10.1093/mnras/stw2699}

\bibitem[{{Veras} {et~al.}(2018){Veras}, {Georgakarakos}, {G{\"a}nsicke}, \&
  {Dobbs-Dixon}}]{Veras2018MNRAS.481.2180V}
{Veras}, D., {Georgakarakos}, N., {G{\"a}nsicke}, B.~T., \& {Dobbs-Dixon}, I.
  2018, \mnras, 481, 2180, \dodoi{10.1093/mnras/sty2409}

\bibitem[{{Veras} {et~al.}(2014){Veras}, {Leinhardt}, {Bonsor}, \&
  {G{\"a}nsicke}}]{Veras2014MNRAS}
{Veras}, D., {Leinhardt}, Z.~M., {Bonsor}, A., \& {G{\"a}nsicke}, B.~T. 2014,
  \mnras, 445, 2244, \dodoi{10.1093/mnras/stu1871}

\bibitem[{{Veras} {et~al.}(2015{\natexlab{b}}){Veras}, {Leinhardt}, {Eggl}, \&
  {G{\"a}nsicke}}]{Veras2015MNRAS.451.3453V}
{Veras}, D., {Leinhardt}, Z.~M., {Eggl}, S., \& {G{\"a}nsicke}, B.~T.
  2015{\natexlab{b}}, \mnras, 451, 3453, \dodoi{10.1093/mnras/stv1195}

\bibitem[{{Veras} {et~al.}(2024){Veras}, {Mustill}, \&
  {Bonsor}}]{Veras2024RvMG}
{Veras}, D., {Mustill}, A.~J., \& {Bonsor}, A. 2024, Reviews in Mineralogy and
  Geochemistry, 90, 141, \dodoi{10.2138/rmg.2024.90.05}

\bibitem[{{Veras} {et~al.}(2013){Veras}, {Mustill}, {Bonsor}, \&
  {Wyatt}}]{Veras2013MNRAS.431.1686V}
{Veras}, D., {Mustill}, A.~J., {Bonsor}, A., \& {Wyatt}, M.~C. 2013, \mnras,
  431, 1686, \dodoi{10.1093/mnras/stt289}

\bibitem[{{Veras} {et~al.}(2016){Veras}, {Mustill}, {G{\"a}nsicke}, {Redfield},
  {Georgakarakos}, {Bowler}, \& {Lloyd}}]{Veras2016MNRAS.458.3942V}
{Veras}, D., {Mustill}, A.~J., {G{\"a}nsicke}, B.~T., {et~al.} 2016, \mnras,
  458, 3942, \dodoi{10.1093/mnras/stw476}

\bibitem[{{Veras} \& {Rosengren}(2023)}]{Veras2023MNRAS}
{Veras}, D., \& {Rosengren}, A.~J. 2023, \mnras, 519, 6257,
  \dodoi{10.1093/mnras/stad130}

\bibitem[{{Veras} \& {Wolszczan}(2019)}]{Veras2019MNRAS.488..153V}
{Veras}, D., \& {Wolszczan}, A. 2019, \mnras, 488, 153,
  \dodoi{10.1093/mnras/stz1721}

\bibitem[{{Veras} {et~al.}(2019){Veras}, {Efroimsky}, {Makarov}, {Bou{\'e}},
  {Wolthoff}, {Reffert}, {Quirrenbach}, {Tremblay}, \&
  {G{\"a}nsicke}}]{Veras2019MNRAS.486.3831V}
{Veras}, D., {Efroimsky}, M., {Makarov}, V.~V., {et~al.} 2019, \mnras, 486,
  3831, \dodoi{10.1093/mnras/stz965}

\bibitem[{{Villaver} \& {Livio}(2009)}]{Villaver2009ApJ}
{Villaver}, E., \& {Livio}, M. 2009, \apjl, 705, L81,
  \dodoi{10.1088/0004-637X/705/1/L81}

\bibitem[{{Villaver} {et~al.}(2014){Villaver}, {Livio}, {Mustill}, \&
  {Siess}}]{Villaver2014ApJ}
{Villaver}, E., {Livio}, M., {Mustill}, A.~J., \& {Siess}, L. 2014, \apj, 794,
  3, \dodoi{10.1088/0004-637X/794/1/3}

\bibitem[{{Virtanen} {et~al.}(2020){Virtanen}, {Gommers}, {Oliphant},
  {Haberland}, {Reddy}, {Cournapeau}, {Burovski}, {Peterson}, {Weckesser},
  {Bright}, {van der Walt}, {Brett}, {Wilson}, {Millman}, {Mayorov}, {Nelson},
  {Jones}, {Kern}, {Larson}, {Carey}, {Polat}, {Feng}, {Moore}, {VanderPlas},
  {Laxalde}, {Perktold}, {Cimrman}, {Henriksen}, {Quintero}, {Harris},
  {Archibald}, {Ribeiro}, {Pedregosa}, {van Mulbregt}, \& {SciPy 1. 0
  Contributors}}]{Virtanen2020}
{Virtanen}, P., {Gommers}, R., {Oliphant}, T.~E., {et~al.} 2020, Nature
  Methods, 17, 261, \dodoi{10.1038/s41592-019-0686-2}

\bibitem[{{Wyatt} {et~al.}(2011){Wyatt}, {Clarke}, \& {Booth}}]{Wyatt2011CeMDA}
{Wyatt}, M.~C., {Clarke}, C.~J., \& {Booth}, M. 2011, Celestial Mechanics and
  Dynamical Astronomy, 111, 1, \dodoi{10.1007/s10569-011-9345-3}

\bibitem[{{Wyatt} {et~al.}(2014){Wyatt}, {Farihi}, {Pringle}, \&
  {Bonsor}}]{Wyatt2014MNRAS}
{Wyatt}, M.~C., {Farihi}, J., {Pringle}, J.~E., \& {Bonsor}, A. 2014, \mnras,
  439, 3371, \dodoi{10.1093/mnras/stu183}

\bibitem[{{Wyatt} {et~al.}(2007){Wyatt}, {Smith}, {Greaves}, {Beichman},
  {Bryden}, \& {Lisse}}]{Wyatt2007ApJ}
{Wyatt}, M.~C., {Smith}, R., {Greaves}, J.~S., {et~al.} 2007, \apj, 658, 569,
  \dodoi{10.1086/510999}

\bibitem[{{Xu} {et~al.}(2016){Xu}, {Jura}, {Dufour}, \&
  {Zuckerman}}]{Xu2016ApJl}
{Xu}, S., {Jura}, M., {Dufour}, P., \& {Zuckerman}, B. 2016, \apjl, 816, L22,
  \dodoi{10.3847/2041-8205/816/2/L22}

\bibitem[{{Zanazzi} \& {Ogilvie}(2020)}]{Zanazzi2020MNRAS}
{Zanazzi}, J.~J., \& {Ogilvie}, G.~I. 2020, \mnras, 499, 5562,
  \dodoi{10.1093/mnras/staa3127}

\bibitem[{{Zhang} {et~al.}(2023){Zhang}, {Naoz}, \& {Will}}]{Zhang2023ApJ}
{Zhang}, E., {Naoz}, S., \& {Will}, C.~M. 2023, arXiv e-prints,
  arXiv:2301.08271, \dodoi{10.48550/arXiv.2301.08271}

\bibitem[{{Zhang} {et~al.}(2021){Zhang}, {Liu}, \& {Lin}}]{Zhang2021ApJ}
{Zhang}, Y., {Liu}, S.-F., \& {Lin}, D. N.~C. 2021, \apj, 915, 91,
  \dodoi{10.3847/1538-4357/ac00ae}

\bibitem[{{Zuckerman} {et~al.}(2003){Zuckerman}, {Koester}, {Reid}, \&
  {H{\"u}nsch}}]{Zuckerman2003ApJ}
{Zuckerman}, B., {Koester}, D., {Reid}, I.~N., \& {H{\"u}nsch}, M. 2003, \apj,
  596, 477, \dodoi{10.1086/377492}

\end{thebibliography}

\end{document}